# Accessibility of Cortical Regions to Focal TES: Dependence on Spatial Position, Safety, and Practical Constraints

Guilherme Bicalho Saturnino[1,2], Hartwig Roman Siebner[1,3,4], Axel Thielscher [1,2*], Kristoffer Hougaard Madsen [1,5*]

[*] shared last authorship

**Institutions:**

[1] Danish Research Centre for Magnetic Resonance, Centre for Functional and Diagnostic Imaging and Research, Copenhagen University Hospital Hvidovre, Denmark.

[2] Department of Health Technology, Technical University of Denmark, Kgs Lyngby, Denmark.

[3] Department of Neurology, Copenhagen University Hospital Bispebjerg, Copenhagen, Denmark

[4] Institute for Clinical Medicine, Faculty of Medical and Health Sciences, University of Copenhagen, Copenhagen, Denmark

[5] Department of Applied Mathematics and Computer Science, Technical University of Denmark, Kgs. Lyngby, Denmark

**Corresponding author**

Dr. Axel Thielscher

Danish Research Centre for Magnetic Resonance,

Centre for Functional and Diagnostic Imaging and Research

Copenhagen University Hospital Hvidovre

DK-2650 Hvidovre, Denmark

Email: axelt@drcmr.dk; Phone #: +45 38623326

# ABSTRACT

Transcranial electric stimulation (TES) can modulate intrinsic neural activity in the brain by injecting weak currents through electrodes attached to the scalp. TES has been widely used as a neuroscience tool to investigate how behavioural and physiological variables of brain function are modulated by electric stimulation of specific brain regions. For an unambiguous interpretation of TES experiments, it is important that the electric fields can be steered towards one or several brain regions-of-interest. However, the conductive proprieties of the human head impose inherent physical limitations on how focal the electric fields in the brain produced by multi-electrode TES can be. As a rule of thumb, it is not feasible to selectively target deep brain areas with TES, although focusing the field in some specific deeper locations might be possible due to favourable conductive properties in the surrounding tissue. In the present study, we first propose a computationally efficient method for the automatic determination of electrode placements and stimulation intensities to optimally affect a given target position. We provide a computationally efficient and robust implementation of the optimization procedure that is able to adhere to safety constraints, while explicitly controlling both the number of active electrodes and the angular deviation of the field in the target area relative to the desired field direction. Leveraging the high computational efficiency of our method, we systematically assess the achievable focality of multi-electrode TES for all cortex positions, thereby investigating the dependence on the chosen constraints. Our results provide comprehensive insight into the limitations regarding the achievable TES dose and focality that are imposed by the biophysical constraints and the safety considerations of TES.

**Keywords:**

Transcranial Electric Stimulation, Optimization, Electric field simulations, Mapping, Focality

# INTRODUCTION

Transcranial Electric Stimulation (TES) is a non-invasive brain stimulation method which aims to facilitate or inhibit neural activity by means of weak currents (usually ≤ 2mA) applied though scalp electrodes. The applied current waveform ranges from direct current (TDCS), alternating currents (TACS) to random noise (TRNS). In the last few years, TES has become a widely used tool for neuromodulation in neuroscience research (Parkin et al., 2015) and clinical applications (Nitsche et al., 2009). However, TES still suffers from a large inter-subject variability (Horvath et al., 2015, 2014; Parkin et al., 2015). One important source of variability is related to the conductive proprieties of the human head. These properties cause the electric fields generated in the brain to exhibit complex, often non-obvious patterns, that depend on the individual anatomy. Therefore, intuitive rule-of-thumb approaches may produce unexpected stimulation patterns and might even miss the intended stimulation target (Miranda et al., 2013; Saturnino et al., 2015). Additionally, these montages stimulate large areas, which makes it difficult to attribute an experimental outcome to the stimulation of a particular brain region. In this context, pseudo-monopolar montages (i.e., ring montages or 4x1 montages) have been recently introduced in order to produce more focal stimulation effects than the classical bi-polar two-electrode montages using a single anode and cathode (Heise et al., 2016; Kuo et al., 2013). In addition, modelling the electric fields might help to reduce the uncertainty of the affected brain areas, allowing more careful planning of individualized electrode montages for targeting a given area of interest. However, manually optimizing montages based on electric field models may imply a lengthy iterative procedure of trial-and-error, especially if one is to fully leverage the potential of multi-channel TES setups. Therefore, methods have been proposed to automatically calculate optimal electrode positions for TES with the aim of targeting given brain regions in an individualized fashion (Dmochowski et al., 2011; Guler et al., 2016a; Park et al., 2011; Ruffini et al., 2014; Sadleir et al., 2012; Wagner et al., 2015).

Determining optimal electrode positions and the corresponding current intensities is a non-linear and non-convex optimization problem. By using a discrete grid of putative electrode positions (e.g., based on the EEG 10-20 system) rather than representing the electrode positions as continuous variables, the problem can be relaxed such that it is convex in certain cases. To the best of our knowledge, the first studies that propose automated methods to optimize multi-channel electrode montages in TES were presented by Dmochowski et al (2011) and Park et al (2011). Since then, several additional methods have been published (Sadleir et al., 2012; Ruffini et al., 2014; Wagner et al., 2015; Guler et al., 2016a). These methods share the same basic approach to employ grids of fixed electrode positions, but they differ, among others, in the way the optimization problems are set-up and how they account for the constraints set by the safety limits for TES and for other practical aspects such as a limited number of available stimulation channels. For example, in the framework proposed by Dmochowski et al. (2011), optimization of the electrode montage is performed either by solving a least squares problem or a linearly constrained quadratic problem, similar to what we propose in the current work. Both settings can be constrained to account for the safety limits. This approach is relatively simple and has several practical advantages over more complex approaches in terms of reliability and speed, but the limitation to a maximal number of channels is not tackled. Guler et al. (2016b) and Ruffini et al. (2014) proposed extended methods based on the branch-and-bound algorithm (Boyd and Mattingley, 2007) and a genetic algorithm, respectively, to overcome this problem, but in both cases at the cost of strongly increased computation times.

In the present study, we propose a novel optimization approach for multi-electrode TES, which can account for the TES safety limits and limited numbers of stimulation channels in an exact and optimal way while offering high computational efficiency. The optimization approach takes practically meaningful parameters as input, such as the target location, the desired electric field at the target, the safety limits governing the maximally injected current strengths and the maximum number of available channels to fully define the problem. We also introduce a new algorithm for constraining the number of active electrodes, where we reduce the time complexity from a few hours, as previously reported (Guler et al., 2016b; Ruffini et al., 2014), to a less than one minute while ensuring optimality.

In addition to characterizing features and performance of the new optimization approach, we use it to systematically map the achievable targeting accuracy and focality across the complete cortex in line with the procedure proposed in (Dmochowski et al., 2017; Huang and Parra, 2019). Specifically, we perform a step-by-step investigation of how the safety limits imposed on the electrode currents, the limited numbers of channels available in TES systems, the selected electric field strength in the target and the amount of alignment of the electric field with the desired field direction affect the quality of the optimized electric fields.

## MATERIALS AND METHODS

### Theoretical Framework

TES applies currents at relatively low frequencies (<10 kHz) at which the head tissues are predominantly resistive and wave effects do not occur (Plonsey and Heppner, 1967), so that the electric fields can be well described in a quasi-static regime. In this case, the field in the human head is governed by the Laplace equation, meaning that the total electric field $\vec{E}(\vec{p})$ created by current injections through multiple (say $n$) electrodes can be described as the sum of the fields $\vec{E}_i(\vec{p})$ caused by the current flows between single $n$-1 electrodes, and an arbitrarily selected common reference electrode (Dmochowski et al., 2011):

$$\vec{E}(\vec{p}) = \sum_{i=1}^{n-1} \vec{E}_i(\vec{p}) \ . \tag{1}$$

Here, $\vec{p}$ denotes a position in the head, and $\vec{E}_i$ are the fields created by the current flow between the $i$-th electrode and the reference electrode. Also, the electric fields scale linearly with the strength of the injected currents, so that the total field can be expressed as a linear superposition of the fields created by unit currents through the electrodes

$$\vec{E}(\vec{p}) = \sum_{i=1}^{n-1} \vec{E}_i(\vec{p}) = \sum_{i=1}^{n-1} \vec{E}_i(\vec{p}, I_i = 1)\, x_i, \tag{2}$$

with $x_i$ being the current injected into the $i$-th electrode, and vector $\vec{E}_i(\vec{p}, I_i = 1)$ the electric field caused by a unit current flowing between the $i$-th electrode and the reference electrode. When assessing $\vec{E}_i(\vec{p}, I_i = 1)$ at a set of $m$ positions in the head, the result can be compactly represented as a vector of length $3m$:

$$\boldsymbol{e}_i = \begin{bmatrix} e_{i,1}^x \\ e_{i,2}^x \\ \vdots \\ e_{i,m}^x \\ e_{i,1}^y \\ \vdots \\ e_{i,m}^z \end{bmatrix}. \tag{3}$$

If we also represent the currents injected in the $n$-1 electrodes as vector $\boldsymbol{x}_a = [x_1\ x_2\ \dots\ x_{n-1}]^T$, the resulting electric field $\boldsymbol{e}$ can be determined by a simple matrix multiplication:

$$\boldsymbol{e} = \boldsymbol{A}_a\, \boldsymbol{x}_a, \tag{4}$$

where

$$\boldsymbol{A}_a = [\boldsymbol{e}_1\ \boldsymbol{e}_2\ \dots\ \boldsymbol{e}_{n-1}]. \tag{5}$$

The matrix $\boldsymbol{A}_a$ has the size of $3m$ x $(n\text{-}1)$ and is referred to as the lead-field matrix (Dmochowski et al., 2011). This matrix is also widely used in EEG source reconstruction, and can be interpreted as an electrical model of the head (Dmochowski et al., 2017). Here, we apply an extended lead-field matrix that includes the reference electrode, as this resulted in a better stability and simpler implementation of the optimization procedure (for further details see below). This is done by introducing a $n\ x\ n-1$ matrix $\boldsymbol{P}$

$$\boldsymbol{P} = \begin{bmatrix} \boldsymbol{I}_{n-1\ x\ n-1} \\ -\boldsymbol{1}_{n-1}^T \end{bmatrix}, \tag{6}$$

where vector $\mathbf{1}_{n-1}^T = [1 \quad 1 \quad \cdots \quad 1]_{1 \times n-1}$ and the matrix $\mathbf{I}_{n-1 \times n-1}$ is the $(n-1) \times (n-1)$ identity matrix. Applying $\mathbf{P}$ and its pseudo inverse $\mathbf{P}^\dagger$ to Eq. 5, we obtain

$$\mathbf{e} = (\mathbf{A}_a \mathbf{P}^\dagger)(\mathbf{P} \mathbf{x}_a), \tag{7}$$

$$\mathbf{e} = \mathbf{A} \mathbf{x}, \tag{8}$$

where $\mathbf{A}$ is the extended lead-field matrix of size $3m \times n$, and the vector $\mathbf{x} = [x_1 \, x_2 \, \ldots \, x_n]^T$ includes the currents injected in all $n$ electrodes, including the reference electrode. Column $i$ of $\mathbf{A}$ corresponds to the electric field which is created by injecting a unit current in the $i$-th electrode, while extracting currents of strength $1/(n-1)$ at the remaining $n-1$ electrodes. After applying the transformation, the system in Eq. 8 becomes ill-posed. That is, it admits many values of $\mathbf{x}$ that result in the same electric field $\mathbf{e}$. However, the only physically plausible solution can be retrieved by imposing Kirchhoff's current law, i.e. that the sum of all incoming and outgoing currents through the electrodes must be zero:

$$\sum_{i=1}^{n} x_i = \mathbf{1}_n^T \mathbf{x} = 0. \tag{9}$$

The matrix-vector product formulation for the electric field (Eq. 8) allows the evaluation of some quantities of interest that are relevant for the optimization procedure described further down:

*1) Mean electric field component in a target region along a specific direction*

In continuous form, the mean field component $\langle \vec{E} \cdot \hat{n} \rangle_{\Omega_\tau}$ in a target volume or area $\Omega_\tau$ and along a target direction $\hat{n}$ is given by

$$\langle \vec{E} \cdot \hat{n} \rangle_{\Omega_\tau} = \frac{1}{G_\tau} \int_{\Omega_\tau} \vec{E} \cdot \hat{n} \, dG, \tag{10}$$

where $G_\tau$ is the total volume or area (depending on the geometry under consideration) of the region $\Omega_\tau$. For a discretized head model, we sample the electric fields at a set of positions, each position representing the electric field in a small volume or area $g_i$. We can therefore write the mean field component at the target area in discrete form as

$$\langle \vec{E} \cdot \hat{n} \rangle_{\Omega_\tau} = \frac{\mathbf{N}_\tau \mathbf{A} \mathbf{x}}{\mathbf{1}^T \mathbf{g}_\tau}, \tag{11}$$

where $\mathbf{g}_\tau$ is a vector of volumes or areas of the positions in the mesh if it is in the target region $\Omega_\tau$, and zero otherwise. The matrix $\mathbf{N}_\tau$ is defined as

$$\mathbf{N}_\tau = [n^x \mathbf{g}_\tau^T \quad n^y \mathbf{g}_\tau^T \quad n^y \mathbf{g}_\tau^T], \tag{12}$$

where $n^x$, $n^y$ and $n^z$ are the x, y and z components of the target direction $\hat{n}$. We can simplify Eq. 11 so that the mean electric field component is represented as a vector-vector multiplication

$$\langle \vec{E} \cdot \hat{n} \rangle_{\Omega_\tau} = \boldsymbol{l}^{\mathrm{T}} \boldsymbol{x}, \tag{13}$$

where

$$\boldsymbol{l}^{\mathrm{T}} = \frac{N_\tau \boldsymbol{A}}{\boldsymbol{1}^{\mathrm{T}} \boldsymbol{g}_\tau}. \tag{14}$$

*2) Total Energy Delivered*

The total energy stored in an electric field in a volume $\Omega_0$ is given by (Griffiths, 1999)

$$U = \frac{\varepsilon_o}{2} \int_{\Omega_0} |\vec{E}(x)|^2 \, dV. \tag{15}$$

Here, we define a quantity proportional to the energy. In the discretized space, the integral in Eq. 15 can be calculated in the discretized head model as

$$\int_{\Omega_0} |\vec{E}(x)|^2 \, dV = \boldsymbol{e}^{\mathrm{T}} \begin{bmatrix} \boldsymbol{G}_0 & 0 & 0 \\ 0 & \boldsymbol{G}_0 & 0 \\ 0 & 0 & \boldsymbol{G}_0 \end{bmatrix} \boldsymbol{e}, \tag{16}$$

where $\boldsymbol{G}_0$ is a diagonal matrix with element area or volumes, if the element is in $\Omega_0$, and zero elsewhere. Using the relation in Eq. 8, we can write

$$\int_{\Omega_0} |\vec{E}(x)|^2 = \boldsymbol{x}^{\mathrm{T}} \boldsymbol{Q} \boldsymbol{x}, \tag{17}$$

where

$$\boldsymbol{Q} = \boldsymbol{A}^{\mathrm{T}} \begin{bmatrix} \boldsymbol{G}_0 & 0 & 0 \\ 0 & \boldsymbol{G}_0 & 0 \\ 0 & 0 & \boldsymbol{G}_0 \end{bmatrix} \boldsymbol{A}. \tag{18}$$

3) Angle between mean electric field and target direction

This quantity can be defined in continuous form, over a target area $\Omega_\tau$ as

$$\theta = \arctan\left( \frac{\sqrt{\frac{1}{G_\tau^2} \int_{\Omega_\tau} (\vec{E} \cdot \hat{n}_{t1})^2 + (\vec{E} \cdot \hat{n}_{t2})^2 \, dG}}{\langle \vec{E} \cdot \hat{n} \rangle_{\Omega_\tau}} \right), \tag{19}$$

where $\hat{n}_{t1}$ and $\hat{n}_{t2}$ are two mutually orthogonal directions, which are also orthogonal to the target direction $\hat{n}$. The discrete form can be written as

$$\theta = \arctan\left(\frac{\sqrt{x^T Q_{\tan} x}}{l^T x}\right), \tag{20}$$

where $Q_{\tan}$ is given by

$$Q_{\tan} = \frac{1}{G_\tau^2} A^T \left( \begin{bmatrix} n_{t1}^x g_\tau & 0 & 0 \\ 0 & n_{t1}^y g_\tau & 0 \\ 0 & 0 & n_{t1}^z g_\tau \end{bmatrix} + \begin{bmatrix} n_{t2}^x g_\tau & 0 & 0 \\ 0 & n_{t2}^y g_\tau & 0 \\ 0 & 0 & n_{t2}^z g_\tau \end{bmatrix} \right) A. \tag{21}$$

## Forward Simulations of the TES Electric Fields

We used the SimNIBS 2.1 software package (Thielscher et al., 2015) to calculate the electric fields that form the columns of the lead-field matrix $A_a$, which we then transformed into the extended lead-field matrix $A$ as stated above. SimNIBS utilizes the Finite Element Method (FEM) with first order tetrahedral elements to calculate electric potentials $\varphi$ by solving the homogeneous Laplace equation

$$\nabla \cdot (\sigma \nabla \varphi) = 0, \tag{22}$$

where $\sigma$ is the ohmic conductivity of the medium. Simulations were performed by applying Dirichlet boundary conditions at the outer boundaries of the $i^{th}$ electrode as well as the return electrode. The electric field

$$\vec{E} = -\nabla \varphi, \tag{23}$$

and the current density

$$\vec{J} = \sigma \vec{E}, \tag{24}$$

were then calculated and the fields were re-scaled to ensure a unit current flow between the electrodes.

The simulations were performed using a head model of a healthy volunteer (Nielsen et al., 2018), created based on existing T1- and T2-weighted magnetic resonance images. Details on image acquisition can be found in (Windhoff et al., 2013). The head model has about $4 \times 10^6$ tetrahedra and $7 \times 10^5$ nodes and was automatically created using the *headreco* routine of SimNIBS 2.1. Conductivities of all tissues were assumed to be isotropic, and the values were assigned to 0.126 S/m (WM), 0.275 S/m (GM), 1.654 S/m (CSF), 0.025 S/m (spongy bone), 0.008 S/m (compact bone), 0.50 S/m (vitreous bodies), 0.465 S/m (scalp), 1 S/m (electrodes) (Saturnino et al., 2015). Electrodes were modelled as small cylinders (1 cm diameter, 2 mm thickness) with homogeneous conductivity, corresponding approximately to electrode gel. Figure 1A shows a sagittal view of the head model with its different tissues.

For optimizing the TES electric fields, we used a fixed electrode grid, shown in Figure 1B. This grid has 288 electrodes, which were positioned semi-automatically to cover the head, neck and face. Additionally, we also created grids with 102, 190, and 381 electrode positions to assess how the field in the cortex depends on density of the electrode grid (Suppl. Figure S1).

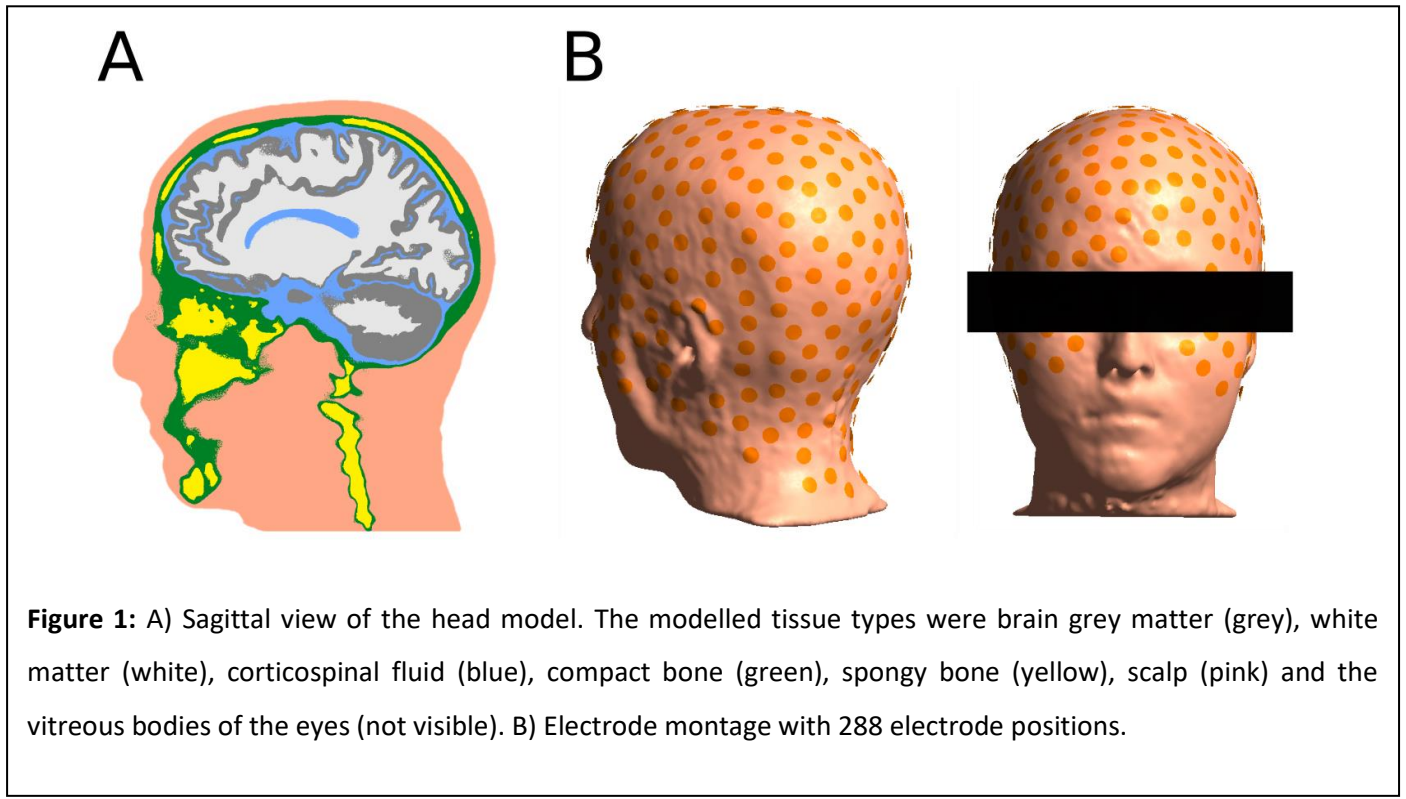

**Figure 1:** A) Sagittal view of the head model. The modelled tissue types were brain grey matter (grey), white matter (white), corticospinal fluid (blue), compact bone (green), spongy bone (yellow), scalp (pink) and the vitreous bodies of the eyes (not visible). B) Electrode montage with 288 electrode positions.

## Optimization of Electrode Currents

When optimizing a TES montage, one might be interested in maximizing the strength of the field at the target. Alternatively, one might want to obtain a focal field while at the same time ensuring that a certain strength of the field is maintained at the target or even at multiple targets simultaneously. In addition to controlling its strength, one might also aim to control the angle of the electric field relative to the orientation of the cortical sheet at the target precisely. In practice, one is often faced with the need to obtain the best possible solution for a limited number of available stimulation channels. To tackle this large variety of optimizations and constraints, we formulated ten different optimization problems, shown in Table 1. The problems are described in detail in the sections below.

| Problem | Objective | Constraints | Problem Class |
|---|---|---|---|
| 1 | Maximize the focality (minimize total energy) | • Maintain the desired electric field in the target | QP |
| 2 | Maximize the field in the target | • Obey safety constraints (total and per-electrode current injected) | LP |
| 3 | Maximize the focality | • Maintain the desired electric field in the target<br>• Obey safety constraints | QP |
| 4 | Maximize the focality | • Maintain the desired electric field in the target<br>• Obey safety constraints<br>• Control the number of active electrodes | Combinatorial |
| 5 | Maximize the field in the target | • Obey safety constraints<br>• Control the field angle in the target | QCLP |
| 6 | Maximize the focality | • Maintain the desired electric field in the target<br>• Obey to the safety constraints<br>• Control the field angle in the target. | QCQP |
| 7 | Maximize the focality | • Maintain the desired electric field in the target<br>• Obey safety constraints<br>• Control the field angle in the target<br>• Control the number of active electrodes. | Combinatorial |
| 8 | Maximize the focality | • Maintain the desired electric fields in *multiple* targets<br>• Obey safety constraints | QP |
| 9 | Maximize the field in *multiple* targets | • Do not exceed a maximum electric field in each target<br>• Obey safety constraints | LP |
| 10 | Maximize the focality | • Maintain the desired electric fields in multiple targets<br>• Obey safety constraints<br>• Control the number of active electrodes. | Combinatorial |

**Table 1:** Optimization problems explored in this paper. The objective refers to what the optimization problem aims to achieve, and the constraints define which requirements the solution must obey. The problem class is a mathematical classification of the problem, and determines the methods, which can be used to effectively solve the problem.

1) Without Safety Constraints

In the simplest setting, we aim to create a given electric field in a target area and direction, while avoiding the other brain areas as well as possible, disregarding the safety constraints that limit how much current one can inject. We formulated this optimization as Problem 1:

$$\begin{aligned} \text{minimize} \quad & \mathbf{x}^\mathrm{T} \mathbf{Q} \mathbf{x} & \text{(P1.1)} \\ \text{subject to} \quad & \mathbf{l}^\mathrm{T} \mathbf{x} = t & \text{(P1.2)} \\ & \mathbf{1}^\mathrm{T} \mathbf{x} = 0 & \text{(P1.2)} \end{aligned}$$

**Problem 1**: Maximization of focality given a target electric field, with no safety constraints. The electrode currents, $\mathbf{x}$, is the optimization variable.

The objective in this optimization problem (Eq. P1.1) is to minimize the energy (Eq. 17). However, we also want to ensure to reach a certain value $t$ for the electric field at our target region and direction (Eq. P1.2, Eq. 13). This is similar to the Linearly constrained minimum variance (LCMV) algorithm proposed in (Dmochowski et al., 2011), with the key difference that the constraint (Eq. P1.2) is to reach the a given average field component in the target region, instead of exactly matching all electric field component in each target location. This means that we have effectively less equality constraints, leaving more degrees of freedom to the solutions. However, this also means that we do not have the same precise control of the field alignment in the target region as (Dmochowski et al., 2011). To reach the objective, we can manipulate the vector of injected currents $\mathbf{x}$, but we must obey Kirchhoff's current law (Eq. P1.2, Eq. 9). The choices for the regions $\Omega_0$, $\Omega_t$ and of the direction $\hat{n}$ are implicit in the matrices $\mathbf{Q}$ and $\mathbf{l}$. Problem 1 is an equality-constrained quadratic programming (QP) problem, and due to convexity and its simplicity, the solution that can be computed directly and efficiently (Boyd and Vandenberghe, 2004).

It is interesting to note that the matrix $\mathbf{Q}$ is $n \times n$ and $\mathbf{l}^\mathrm{T}$ is $1 \times n$, while the leadfield matrix $\mathbf{A}$ is $3m \times n$. In TES optimization, $n$ is the number of electrodes (in the range of 32 to 500) while $m$ is the number of positions in the region of interest (the brain or the whole head), typically between $10^5$ and $10^6$. Therefore, this formulation reduces the size of the problem, which is closely linked to computation time, by 3 to 4 orders of magnitude when comparing with formulations that use the leadfield matrix directly such as the genetic algorithm proposed in (Ruffini et al., 2014), or the formulation proposed by (Wagner et al., 2015).

2) With Safety Constraints

The most common safety constraint in TES is to limit the total injected current to a given value $I_{\text{tot}}$. As the total inflowing current is equal to the total outflowing current (as per Kirchhoff's current law), we can write this constraint as

$$\sum_{i=1}^{n} |x_i| \leq 2I_{\text{tot}}. \qquad (25)$$

Which corresponds to

$$\|x\|_1 \leq 2I_{\text{tot}}, \qquad (26)$$

where $\|\cdot\|_1$ denotes the L1-norm. When using small electrodes, it can also be of interest to limit the current flowing through each electrode to a given value $I_{\text{ind}}$ (inflowing or outflowing) to avoid skin irritation, discomfort and heating at individual electrode interfaces. This corresponds to a bound constraint on the current through each electrode

$$-I_{\text{ind}} \leq x_i \leq I_{\text{ind}}, \quad i = 1, 2, \ldots, n. \qquad (27)$$

The simplest optimization which considers the safety constraints (Eq. 26 and Eq. 27) is Problem 2:

$$\begin{aligned} \text{maximize} \quad & \boldsymbol{l}^T \boldsymbol{x} & (P2.1) \\ \text{subject to} \quad & \mathbf{1}^T \boldsymbol{x} = 0 & (P2.2) \\ & \|x\|_1 \leq 2I_{\text{tot}} & (P2.3) \\ & -I_{\text{ind}} \leq x_i \leq I_{\text{ind}}, \quad i = 1, 2, \ldots, n & (P2.4) \end{aligned}$$

**Problem 2**: Maximization of the electric field at a target, including safety constraints.

Here, the objective (Eq. P3.1) is to maximize the field at the target region and direction (Eq. 13) while obeying Kirchhoff's current law (Eq. P2.2, Eq. 9) and the safety constraints. In this problem, we do not consider the focality of the field. This is an instance of a linear programming (LP) problem, similar to the *intensity optimization* problem in (Dmochowski et al., 2011), with the difference being the definition of the matrix $\boldsymbol{l}$ (see the description of Problem 1 for more information). In our implementation, we solved it using the SciPy implementation of the simplex algorithm (Dantzig, 1963). However, there exists a very simple solution to this problem, as one can show that Problem 2 can be solved by simply choosing the largest and the smallest entries in $\boldsymbol{l}$, setting the electrode currents in those electrodes to $I_{\text{ind}}$ and $-I_{\text{ind}}$, and repeating the process until the safety constraint in Eq. P2.3 is reached. This results in $\text{ceil}(2I_{\text{tot}}/I_{\text{ind}})$ active electrodes. A similar procedure has been previously described in (Fernandez-corazza et al., 2019).

To obtain a focal field, we introduce Problem 3:

$$\begin{aligned}
\text{minimize} \quad & \mathbf{x}^\text{T}\mathbf{Q}\mathbf{x} & \text{(P3.1)} \\
\text{subject to} \quad & \mathbf{l}^\text{T}\mathbf{x} = t & \text{(P3.2)} \\
& \mathbf{1}^\text{T}\mathbf{x} = 0 & \text{(P3.3)} \\
& \|\mathbf{x}\|_1 \leq 2I_{\text{tot}} & \text{(P3.4)} \\
& -I_{\text{ind}} \leq x_i \leq I_{\text{ind}}, \quad i = 1, 2, \ldots, n & \text{(P3.5)}
\end{aligned}$$

**Problem 3**: Obtain a focal field given a target electric field, including safety constraints.

Problem 3 similar to the linearly constrained minimum variance (LCMV) approach described in (Dmochowski et al., 2011), but with both the L1 and individual electrode constraints. Again, one key difference is the definition of the equality constraint (Eq. P3.2), which is explained above. This formulation fully specifies the optimization problem with a few and intuitively interpretable parameters to be defined by the user: a target direction, a target region, the desired average field in the target region $t$, the maximal current through each electrode $I_{\text{ind}}$, and the maximum total current injected $I_{\text{tot}}$. Problem 3 can be transformed into a quadratic problem (QP) in standard form (Boyd and Vandenberghe, 2004) and solved with standard QP algorithms. For this study, we implemented the active-set method described in (Gill and Murray, 1978) in Python, using the NumPy and SciPy packages (van der Walt et al., 2011), as in our experience, employing the generic general convex optimizer CVXPY (Diamond and Boyd, 2016) is computationally inefficient and often fails to converge. Our implementation has robustly converged in thousands of different settings with minimal computational effort, as shown in the Results section. In our implementation, if the value for $t$ is chosen too high in relation to the values $I_{\text{tot}}$ and $I_{\text{ind}}$, Problem 3 becomes infeasible and we fall back to solving Problem 2. We chose this approach as it is guaranteed to give the same solution, as one would obtain by successively lowering the target $t$ in Problem 3 until it becomes just feasible.

3) Constraining the number of active electrodes

In practice, a limited number of stimulation channels is available, so that it is important to constrain the number of active electrodes to a number $N$. This corresponds to an additional constraint in the form

$$\|x\|_0 \leq N, \tag{28}$$

where $\|\cdot\|_0$ is the L0-norm (or cardinality), that is, the number of non-zero elements in the vector. As discussed above in the description of Problem 2, the L0-norm of the result obtained when maximizing the intensity at the target (Problem 2) is naturally constrained by the safety constraints. However, applying this constraint to Problem 3, we obtain Problem 4:

$$\begin{aligned}
\text{minimize} \quad & x^T Q x & \text{(P4.1)} \\
\text{subject to} \quad & l^T x = t & \text{(P4.2)} \\
& \mathbf{1}^T x = 0 & \text{(P4.3)} \\
& \|x\|_1 \leq 2 I_{\text{tot}} & \text{(P4.4)} \\
& -I_{\text{ind}} \leq x_i \leq I_{\text{ind}}, \quad i = 1, 2, \ldots, n & \text{(P4.5)} \\
& \|x\|_0 \leq N & \text{(P4.6)}
\end{aligned}$$

**Problem 4**: Obtain a focal field, given a target electric field, including safety constraints and limiting the number of active electrodes.

Due to the new constraint (Eq. P4.6), Problem 4 is non-convex. In fact, this type of problem is of combinatorial complexity (Boyd and Vandenberghe, 2004). Therefore, obtaining the globally optimal solution to the problem in a computationally efficient way can be challenging. Here, we use a Branch-and-Bound method (Boyd and Mattingley, 2007) that stems from the same class of methods as used in (Guler et al., 2016b). Details of the algorithm are described in the supplementary material. Shortly, the Branch-and-Bound algorithm consists of successively partitioning (branching) a given region of the parameter space $Q$, which defines active and inactive electrodes. We define a lower bound $\Phi_{lb}(Q)$ and an upper bound $\Phi_{ub}(Q)$ function that can calculate bounds the optimum value of the objective function (Eq. P4.1) within the parameter space $Q$ (see the supplementary material for mathematical details on the bound functions). Repeated branching of the parameter space and evaluation of the bounds results in narrowing down the parameter space, eventually giving us a solution, which is certified to be close to the global optimum.

For quickly searching the parameter space, we make use of the fact that the constraint on the total injected current $I_{\text{tot}}$ (Eq. 26) involves the sum of the absolute values (the L1-norm) of the vector of current values $x$. This term is well-investigated (Tibshirani, 1996) and is known to induce sparsity of the solutions (i.e., many non-active electrodes). Therefore, the solutions of Problem 3 tend to be sparse. We can also add more L1 constraints as a relaxation of the L0 constraint to obtain tight lower bounds $\Phi_{lb}(Q)$ for the branch-and-Bound algorithm.

In the current paper, we stopped the Branch-and-Bound iterations once we obtained solutions which are certified to be within 10% of the global optima. For a problem with 288 possible electrode positions, but only six active electrodes, we only needed more than 20 steps of the Branch-and-Bound algorithm in about 1% of the positions studied. The optimal value is typically found within the first few iterations, and the remaining iterations are required to certify that the solution found is indeed close to the global solution. This behaviour is expected, as the employed L1-heuristcs gives often good approximations of the L0 norm (Boyd and Mattingley, 2007). For problems where more active electrodes are allowed, the algorithm tends to converge even faster, as the L0 constraint plays a smaller role in limiting the solution.

4) Constraining the angular deviation of the electric field in the target

The TES effects on neural activity depend on the direction of the electric field relative to the neural target structures. Specifically, it is often assumed that TES acts predominantly via the polarization of cortical pyramidal cells, so that it is most effective when the fields are oriented normally to the local cortical surface (Stagg and Nitsche, 2011). It is unclear how sensitive the physiological TES effects are to deviations of the field from a perfect normal orientation. In addition, for many target positions, we may have to trade-off field intensity and the accuracy if we want to reach the optimal direction during optimization. For those reasons, it is helpful to have a flexible control of the acceptable amount of deviation from the ideal orientation during the optimization process. Our optimization scheme gives the possibility to constrain the angle between the mean electric field vector in the target and the selected target direction to a given maximal value $\theta_{\max}$. From Eq. 20, we have that:

$$\arctan\left(\frac{\sqrt{x^T Q_{\tan} x}}{l^T x}\right) \leq \theta_{\max}, \tag{29}$$

$$\frac{\sqrt{x^T Q_{\tan} x}}{l^T x} \leq \tan(\theta_{\max}), \tag{30}$$

$$\frac{x^T Q_{\tan} x}{(l^T x)^2} \leq \tan^2(\theta_{\max}), \qquad l^T x \geq 0. \tag{31}$$

Considered in isolation, this constraint is non-convex. However, in combination with the constraint to reach a mean electric field in the target $l^T x = t$, it can be reformulated as a quadratic constraint in the form

$$x^T Q_{\tan} x \leq \tan^2(\theta_{\max}) t^2. \tag{32}$$

Adding this constraint to Problem 2, we obtain Problem 5, which is a quadratically constrained linear programming (QCLP) problem:

$$\text{maximize} \quad l^T x \quad \quad \text{(P5.1)}$$

$$\text{subject to} \quad \mathbf{1}^T x = 0 \quad \quad \text{(P5.2)}$$

$$\|x\|_1 \leq 2 I_{\text{tot}} \quad \quad \text{(P5.3)}$$

$$-I_{\text{ind}} \leq x_i \leq I_{\text{ind}}, \quad i = 1, 2, \ldots, n \quad \text{(P5.4)}$$

$$x^T Q_{\text{tan}} x \leq \tan^2(\theta_{\text{max}}) \, t^2 \quad \quad \text{(P5.6)}$$

**Problem 5**: Maximization of the electric field at a target, including safety constraints and angle constraint.

Similarly, when also considering the focality as formulated in Problem 3, we obtain Problem 6 which is a quadratically constrained quadratic programming (QCQP) problem.

$$\text{minimize} \quad x^T Q x \quad \quad \text{(P6.1)}$$

$$\text{subject to} \quad l^T x = t \quad \quad \text{(P6.2)}$$

$$\mathbf{1}^T x = 0 \quad \quad \text{(P6.3)}$$

$$\|x\|_1 \leq 2 I_{\text{tot}} \quad \quad \text{(P6.4)}$$

$$-I_{\text{ind}} \leq x_i \leq I_{\text{ind}}, \quad i = 1, 2, \ldots, n \quad \text{(P6.5)}$$

$$x^T Q_{\text{tan}} x \leq \tan^2(\theta_{\text{max}}) \, t^2 \quad \quad \text{(P6.6)}$$

**Problem 6**: Obtain a focal field, given a target electric field, including safety constraints and angle constraints.

In our implementation, we solve Problems 5 and 6 by adding the angle constraint (Eq. P5.6 and Eq. P6.6) to the objective weighted by a variable $\lambda$ ($\lambda \geq 0$), and iterating over values of $\lambda$ to find the smallest value of $\lambda$ where the angle constraint is fulfilled. The angle constraint can also be combined with the constrained on the number of active electrodes, resulting in Problem 7

$$\begin{aligned}
&\text{minimize} && x^T Q x && \text{(P7.1)} \\
&\text{subject to} && l^T x = t && \text{(P7.2)} \\
& && \mathbf{1}^T x = 0 && \text{(P7.3)} \\
& && \|x\|_1 \leq 2 I_{\text{tot}} && \text{(P7.4)} \\
& && -I_{\text{ind}} \leq x_i \leq I_{\text{ind}}, \ i = 1, 2, \ldots, n && \text{(P7.5)} \\
& && x^T Q_{\text{tan}} x \leq \tan^2(\theta_{\max}) t^2 && \text{(P7.6)} \\
& && \|x\|_0 \leq N && \text{(P8.7)}
\end{aligned}$$

**Problem 7**: Obtain a focal field given a target electric field, including safety constraints, angle constraints and a constraint on the maximum number of active electrodes.

To solve this problem, we run a branch-and-bound function at each value of the parameter $\lambda$. The problems 5-7 can also be solved with a conic solver (Boyd and Vandenberghe, 2004).

5) Optimization for multiple targets

We can also use the algorithm to optimize the electric fields in several distant targets at once. The simplest way to do it is to extent the definition of the target area $\Omega_\tau$. However, as the operator $l^T$ calculates the average across the entire target region, it might cause a bias towards one "easier to reach" target over others. We overcome this by defining multiple independent target regions $\Omega_{\tau_1}, \Omega_{\tau_2}, \ldots, \Omega_{\tau_q}$ and enforcing the equality constraint (Eq. 13) for each region individually. With this setting, we can also define different target values $t_1, t_2, \ldots, t_q$ for each target region. By adding multiple targets to Problem 3, we obtain Problem 8.

$$\text{minimize} \quad x^T Q x \tag{P8.1}$$

$$\text{subject to} \quad \begin{bmatrix} l_1^T \\ l_2^T \\ \vdots \\ l_q^T \end{bmatrix} x = \begin{bmatrix} t_1 \\ t_2 \\ \vdots \\ t_q \end{bmatrix} \tag{P8.2}$$

$$\mathbf{1}^T x = 0 \tag{P8.3}$$

$$\|x\|_1 \leq 2 I_{\text{tot}} \tag{P8.4}$$

$$-I_{\text{ind}} \leq x_i \leq I_{\text{ind}}, \quad i = 1, 2, \dots, n \tag{P8.5}$$

**Problem 8**: Obtain a focal electric field with multiple targets.

Problem 8 is still a QP, so it can be solved using the same algorithms as used for Problem 3. In cases where Problem 8 is infeasible because the target field values cannot be reached in at least one of the targets, we instead solve Problem 9

$$\text{maximize} \quad \mathbf{1}^T \begin{bmatrix} l_1^T \\ l_2^T \\ \vdots \\ l_q^T \end{bmatrix} x \tag{P9.1}$$

$$\text{subject to} \quad \begin{bmatrix} l_1^T \\ l_2^T \\ \vdots \\ l_q^T \end{bmatrix} x \leq \begin{bmatrix} t_1 \\ t_2 \\ \vdots \\ t_q \end{bmatrix} \tag{P9.2}$$

$$\mathbf{1}^T x = 0 \tag{P9.3}$$

$$\|x\|_1 \leq 2 I_{\text{tot}} \tag{P9.4}$$

$$-I_{\text{ind}} \leq x_i \leq I_{\text{ind}}, \quad i = 1, 2, \dots, n \tag{P9.5}$$

**Problem 9**: Maximize the electric field at multiple targets

That is, we maximize the electric field across all targets, but ensure that the mean electric field at each target does not exceed the given target value (Eq. P9.2). This prevents one easy-to-hit target to take over the objective function (Eq. P9.1), and thus we obtain a field best approaches the target intensities values $t_i$ but without surpassing it. Finally, we add the constraint to the number of active electrodes, obtaining Problem 10:

$$\text{minimize} \quad x^\text{T} Q x \qquad (P10.1)$$

$$\text{subject to} \quad \begin{bmatrix} l_1^\text{T} \\ l_2^\text{T} \\ \vdots \\ l_q^\text{T} \end{bmatrix} x = \begin{bmatrix} t_1 \\ t_2 \\ \vdots \\ t_q \end{bmatrix} \qquad (P10.2)$$

$$\mathbf{1}^\text{T} x = 0 \qquad (P10.3)$$

$$\|x\|_1 \leq 2 I_\text{tot} \qquad (P10.4)$$

$$-I_\text{ind} \leq x_i \leq I_\text{ind}, \ i = 1, 2, \ldots, n \qquad (P10.5)$$

$$\|x\|_0 \leq N \qquad (P10.7)$$

**Problem 10**: Obtain a focal electric field with multiple targets, with a constrained number of electrodes.

Problems 8-9 can all be solved with the same QP, LP and Branch-and-Bound solvers presented earlier. We can also constrain the angle at each target individually, by adding one angle constraint (Eq. 31) for each target region. This problem will not be addressed in the current work.

## Performance and Error Metrics

We define several metrics to characterize the quality of the electric field distributions achieved by the optimization approaches:

*1) Targeting Error*

Often the maximal electric field will not be in the target node. Therefore, we measure the Euclidian distance between the node where the maximum electric field strength $|\vec{E}|$ occurs and the target position (Dmochowski et al., 2017).

*2) Effective Area*

In order to quantify the achieved focality of the optimized electric field, this metric weights the area of the region $\Omega_0$ by the electric field strength. This metric is closely related to the energy (Eq. 6). To enable direct comparison across conditions, we normalize it by the electric field at the target position:

$$\left|\vec{E}\right|_\text{area} = \frac{\sum_{i=1}^{m} |\vec{E}_i| g_i}{l^\text{T} x}. \qquad (33)$$

Here, $\vec{E}_i$ is the electric field in the $i$-th element and $g_i$ its area.

*3) Area with $|\vec{E}|$ exceeding 50% of the target electric field*

As an alternative index of focality, we use the total area with an electric field strength larger or equal 50% of the target electric field $t$:

$$A_{\text{stim}} = \sum_{i=1}^{m} \begin{cases} g_i, & |\vec{E}_i| \geq 0.5t \\ 0, & \text{else} \end{cases}. \qquad (34)$$

The targeting error, the effective area and the stimulated area serve to quantify different aspects of the quality of the solution. The targeting error assesses how close the strongest stimulated point was to the target, and the focality measures give insight into how much the field could be focused there. Specifically, the effective area closely resembles the energy (Eq. 15) and hence is directly related to the optimization procedure. We therefore expect this metric to increase monotonically as we impose tighter constraints on the optimization. In contrast, the targeting error is not directly assessed in the optimization procedure, therefore the dependence on the constraints is less predictable. In order to enable easier comparisons across the tested conditions, we report differences of the above metrics when appropriate. For the targeting error, we report absolute differences in [mm]. For the Effective Area, we calculate the difference in [%] relative to a reference condition:

$$\Delta(\text{Effective Area}) = 100 \times \frac{\text{Effective Area} - \text{Effective Area}_{\text{ref}}}{\text{Effective Area}_{\text{ref}}} \%. \qquad (35)$$

## Evaluation of Optimization Performance across the Brain

We systematically evaluated the performance of the optimization approach for target positions distributed across the complete cortex. This evaluation was initially performed for the optimization without safety constraints (Problem 1), and then repeated after successively adding constraints for safety (Problems 2 and 3), the number of electrodes (Problem 4) and field angle in the target (Problems 6 and 7). We extracted the electric field values created by the optimized montages in the middle of the grey matter layer, which is where TES is expected to be effective in modulating the membrane potential of pyramidal cells (Stagg and Nitsche, 2011). For that, we first determined the location of the middle grey matter sheet by taking the average of the pial and white matter surfaces of the FreeSurfer segmentation. This surface was then corrected for self-intersections and degenerate triangles, and down-sampled to 20,000 nodes using MeshFix (Attene, 2010). We then interpolated the columns of the lead-field matrix (which correspond to electric fields) into the nodes of this surface mesh using the super-convergent patch recovery (SPR) approach described in (Saturnino et al., 2018). To map the optimal fields across the cortex, we optimized the electric field to target at each mesh node, one at a time, and calculated the field metrics. The procedure is described in Algorithm 1.

---

**Input:** Leadfield $A$, Areas or Volumes $g$, Target Orientations $N$, Optimization problem $\mathcal{P}$

1. **Initialize** empty vectors $\mathbf{v}_T, \mathbf{v}_E$ and $\mathbf{v}_A$ to store the measures described in the *Performance and Error Metrics* section
2. **Calculate** Energy Matrix $Q$ (Eq. 18)
3. **For** each column $i$ in $A$
4.    **Calculate** $l^\mathrm{T}$ and $Q_\mathbf{tan}$ for the row $i$ and direction $N_i$
5.    **Solve** the optimization problem $\mathcal{P}$, obtaining the electrode currents $x$
6.    **Calculate** the electric field $e = Ax$
7.    **Calculate** the performance and error metrics and store them in $\mathbf{v}_T, \mathbf{v}_E$ and $\mathbf{v}_A$
8. **Return** $\mathbf{v}_T, \mathbf{v}_E$ and $\mathbf{v}_A$

**Algorithm 1**: Mapping Procedure

---

This approach is similar to previously published studies (Dmochowski et al., 2017; Huang and Parra, 2019). For the figures shown in the paper, the target direction of the electric field $\hat{n}$ was chosen normal to the local surface orientation. In the supplementary material, complementary results are depicted which show metrics averaged over two optimizations, with orthogonal directions defined in the tangent plane.

## RESULTS

### Comparison to Ad-hoc Montages

To provide intuitive insight into the prospects and limitations of optimized multichannel TES montages, we compared the electric fields obtained by ad-hoc montages versus the fields created by optimized approaches. We started by simulating a "classical" montage to target the left motor cortex using two pad electrodes (upper row of Fig. 2A), which results in a non-focal field distribution. We then derived an optimized montage which generates the same electric field in the sulcal wall of the hand knob (Yousry et al., 1997). For that, we extracted the field caused by the "classical" montage at the centre position of the hand knob (shown in the third column of Fig. 2A) and used it as input (i.e., as target value $t$) for an optimization with the safety constraints (Problem 3, $I_\mathrm{ind} = 1\,\mathrm{mA}$, $I_\mathrm{tot} = 2\,\mathrm{mA}$) and using 288 electrode positions (Figure 1B). It is evident that optimization significantly improved the field focality over the Ad-hoc montage (lower row of Fig. 2A). Next, we compared a centre-surround ring montage, aimed at creating an electric field in the crown of left precentral gyrus forming the hand knob, to the optimized montage that creates the same field in the target position (Fig. 2B; same safety constraints and number of electrode positions as in the example before). While the ring montage already induces a quite focal field, this is still improved by the optimized montage.

Finally, we compared ad-hoc vs. optimized montages for the stimulation of a slightly deeper area in the left cingulate sulcus (Fig. 2C). In order to maximize the field in that region, we intuitively chose the positions of the two pad electrodes for the ad-hoc montage to be on the left and right sides of the head at approx. the same height as the target. The optimized montage achieves substantially better focality, but still causes the strongest fields in extended areas of the brain surface. This example highlights the fact that the TES optimization is fundamentally limited by the underlying physics of the head which acts as ohmic volume conductor. The focality metrics for these montages are shown in Table 2. Again, a clear improvement of all metrics is obtained when comparing with the ad-hoc approaches. It is interesting to note that the optimized montages shown in Figure 2 have only few active electrodes. This behaviour of the optimization approach is further investigated below.

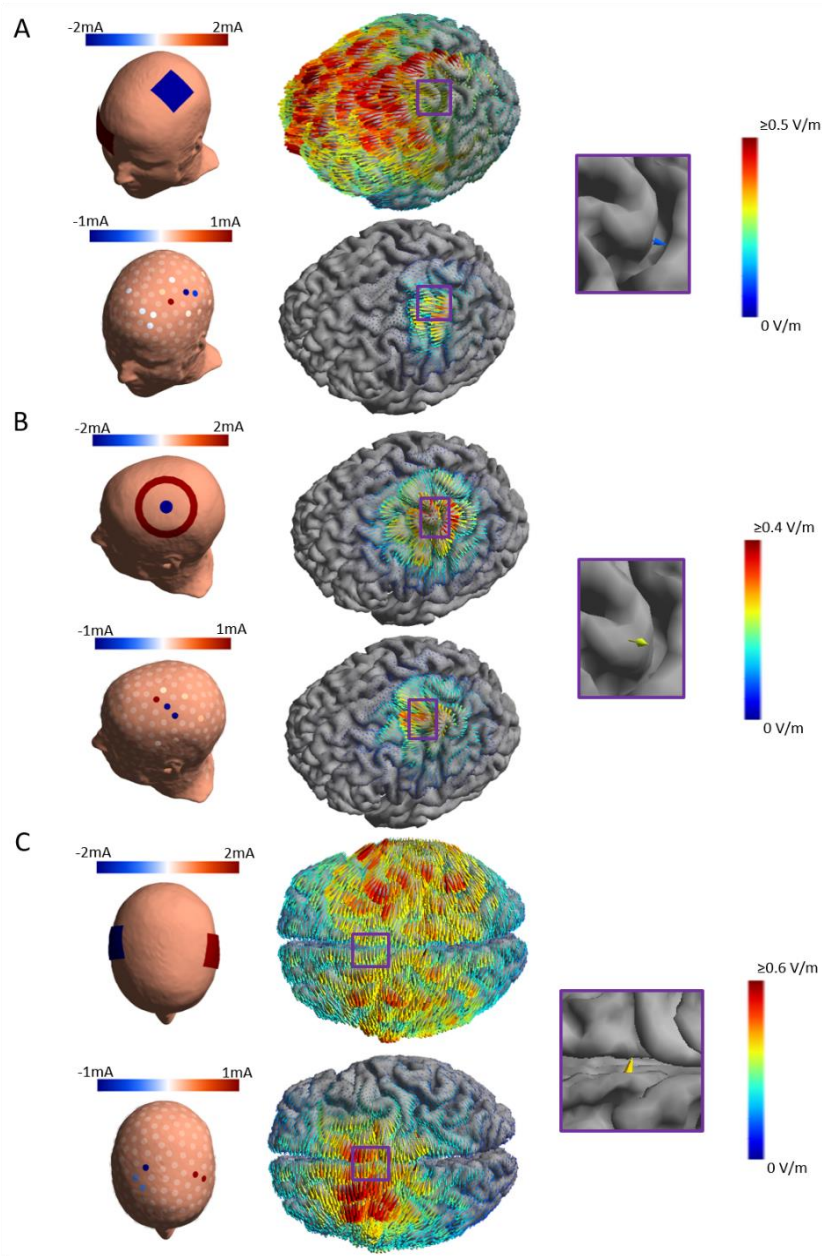

**Figure 2:** Comparison between ad-hoc and optimized multi-electrode montages. A) Upper row: "Classical" montage to stimulate the left motor cortex, here with a 5x5 cm² square electrode over the left hand knob and a 7x5 cm² rectangular "return" electrode over the contralateral supraorbital area. Lower row: Optimized montage that achieves the same field in the sulcal wall in the central part of the hand knob. Current strength in the electrodes is coded as varying levels of red and blue, with inactive electrodes being shown as white and semi-transparent. Third column: Target field at the centre position of the hand knob used as input for the optimization. The target field matches the electric field caused by the "classical" montage at the same position. B) Fields caused by a centre-surround ring montage centred above the left hand knob, and the corresponding optimized multi-electrode montage. The ring montage consists of a circular electrode with 2 cm diameter and ring electrode with 7.5 cm inner and 10 cm outer diameter. C) Stimulation of a position in the left cingulate sulcus. The ad-hoc montage consists of two 5x5 cm² square electrodes positioned laterally at the same height as the target position.

|  |  | Targeting Error (mm) | Area with $\|\vec{E}\| \geq 0.5t$ (cm$^2$) | Effective Area (cm$^2$) |
|---|---|---|---|---|
| **Hand Knob (Sulcal Wall)** | Ad hoc | 84 | 1744 | 3853 |
|  | Optimal | 12 | 238 | 470 |
|  | Difference | -72 mm | -86% | -86% |
| **Hand Knob (Gyral Crown)** | Ad hoc | 13 | 102 | 214 |
|  | Optimal | 4 | 57 | 200 |
|  | Difference | -9 mm | -44% | -7% |
| **Premotor** | Ad hoc | 37 | 1132 | 1048 |
|  | Optimal | 17 | 347 | 638 |
|  | Difference | -20 mm | -69% | -39% |

**Table 2:** Performance metrics for the ad-hoc and optimized montages for the 3 targets shown in Figure 2. The optimized montages consistently have lower targeting errors (distance between the electric field maximum and the desired target position) and better focality. Unsurprisingly, the difference is strongest for the "standard" motor cortex montage (Fig. 2A). Focality was assessed by measuring the total area affected by an electric field of at least 50% of the target electric field, and by calculating the "Effective stimulation area" (i.e., the area weighted by the norm of the electric field, normalized by the electric field at the target; short: "Effective area"). The total area of the middle cortical sheet is $1744$ cm$^2$.

## Maximally Achievable Focality without Safety Constraints

To establish a best-case scenario, we optimized the electric field focality at each of the 20,000 positions in the middle of the modelled grey matter sheet, without considering safety constraints (Problem 1). The target orientations $\hat{n}$ were chosen normal to the surface, and the target electric field was set to $t = 0.3$ V/m. Please notice that the later value is arbitrary, as safety constraints were not considered and the focality metrics are normalized. The latter will thus not change when settings a different target electric field. The metrics for the optimized fields are shown in Figure 3. It is interesting to notice that, even for this best-case scenario without safety constraints, the metrics differ strongly depending on the position of the cortical target. Areas in inferior regions, in the sulci and in medial portions of the brain exhibit increased targeting errors and a worse focality than targets in superficial gyri. Again, this is to be expected, because the currents necessarily need to flow through other brain areas to hit deeper targets. Furthermore, we note that the two area-based metrics (Effective Area and Area with $|\vec{E}| > 0.5t$) have very similar qualitative behaviours, therefore we only report the Effective Area in rest of the study. We opted for this area metric, as it does not require choosing an arbitrary cut-off value.

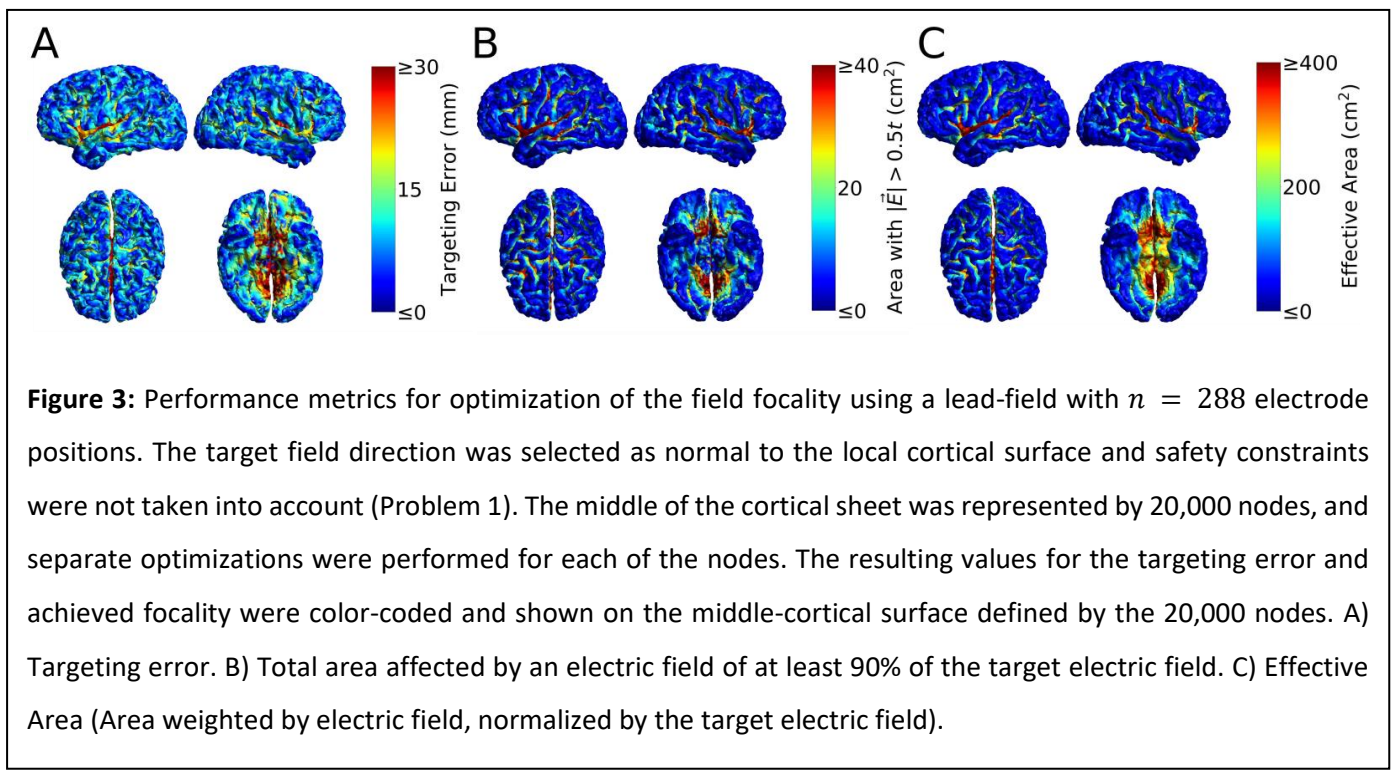

**Figure 3:** Performance metrics for optimization of the field focality using a lead-field with $n = 288$ electrode positions. The target field direction was selected as normal to the local cortical surface and safety constraints were not taken into account (Problem 1). The middle of the cortical sheet was represented by 20,000 nodes, and separate optimizations were performed for each of the nodes. The resulting values for the targeting error and achieved focality were color-coded and shown on the middle-cortical surface defined by the 20,000 nodes. A) Targeting error. B) Total area affected by an electric field of at least 90% of the target electric field. C) Effective Area (Area weighted by electric field, normalized by the target electric field).

The grid of electrode positions is required to have a sufficiently high density in order to minimize the targeting error and to obtain fields that are as focal as possible. We determined an adequate grid density by optimizing the electric field focality (Problem 1) for grids with with $n = 102, 190, 288,$ and $381$ scalp electrodes (Suppl. Fig. S2). Both the targeting accuracy and the focality improve with increasing number of electrodes for up to 288 positions, but improve only marginally further for 381 positions. This is expected, as the conductive proprieties of the head, in particular the low conductivity of the skull coupled with the comparatively high conductivity of CSF, cause the electric fields to spatially disperse and intrinsically limit the achievable spatial resolution (Dmochowski et al., 2012). These findings motivated our choice of using 288 positions in the main part of the study. In Suppl. Fig. S3 and S4 we additionally show focality results for subcortical targets and targets oriented tangentially to the cortical sheet, respectively, where the same patterns can be observed.

## Maximally Achievable Intensity in the Presence of Safety Constraints

In practice, the above best-case scenario cannot be reached due to the requirement that safety constraints are obeyed. Naturally, limiting the current that is injected in the electrodes will also limit the strength of the field that can be achieved in the brain. In a first step, we were thus interested in the maximal strength of the field that can be reached at each cortex position when the current injected in each individual electrode has to stay within $I_{ind} = 1$ mA and the current injected in total has to be within $I_{tot} = 2$ mA. For that, we determined the maximal electric field in the 20,000 cortical positions without aiming to reach a focal stimulation (Problem 2), again using a target orientation normal to the cortical sheet and a grid of 288 electrode positions in the calculations. Expectedly, the maximally achievable field was highest at gyral targets for which it reached 0.6 V/m at some positions (Figure 4). Substantially weaker fields could be achieved in sulcal and deep targets, which, however, still exceeded 0.3 V/m at most positions. As focality was not enforced, both the targeting error and the focality (Effective Area) were far worse than in for the best-case scenario (Figure 3).

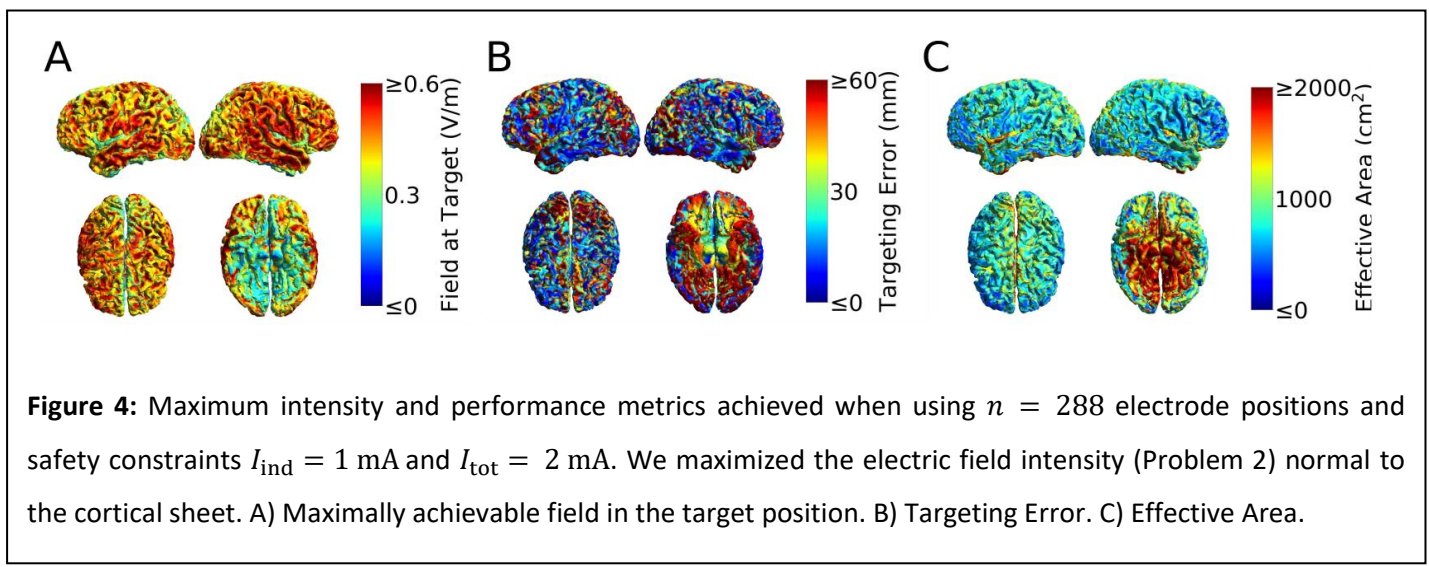

**Figure 4:** Maximum intensity and performance metrics achieved when using $n = 288$ electrode positions and safety constraints $I_{ind} = 1$ mA and $I_{tot} = 2$ mA. We maximized the electric field intensity (Problem 2) normal to the cortical sheet. A) Maximally achievable field in the target position. B) Targeting Error. C) Effective Area.

In Suppl. Fig. S5, we repeated the analysis above for lead-fields with $n = 102, 190, 288$ and $381$ electrodes. The achievable field intensity at the target increased up to 288 electrodes, but was almost unchanged when further increasing the number of electrodes to 381. Suppl. Figs. S6 and S7 show the same analysis for subcortical targets and for targets with tangent orientations to the cortical sheet, respectively, confirming the above results.

## Effect of Safety Constraints on the Optimization of Focality

Compared to the mere maximization of the intensity in the target, a more refined strategy is to make the field as focal as possible while ensuring that a desired, sufficiently high intensity is achieved at the target location. Figure 5 shows the performance of this strategy when obeying safety constraints of $I_{\text{ind}} = 1$ mA and $I_{\text{tot}} = 2$ mA and ensuring a target field strength of $t = 0.2$ V/m. The results depend strongly on the target position, and a high targeting error and low focality dominate for deep targets. Comparison of the results to those obtained for optimization without safety constraints (Figure 3A&C) shows that the targeting error and the effectively stimulated area increase most in inferior and medial brain regions, i.e. in areas that are difficult to target in general. Suppl. Figs. S9 and S10 show the same analysis for subcortical targets and for targets with tangential orientations to the cortical sheet, respectively, confirming the above results.

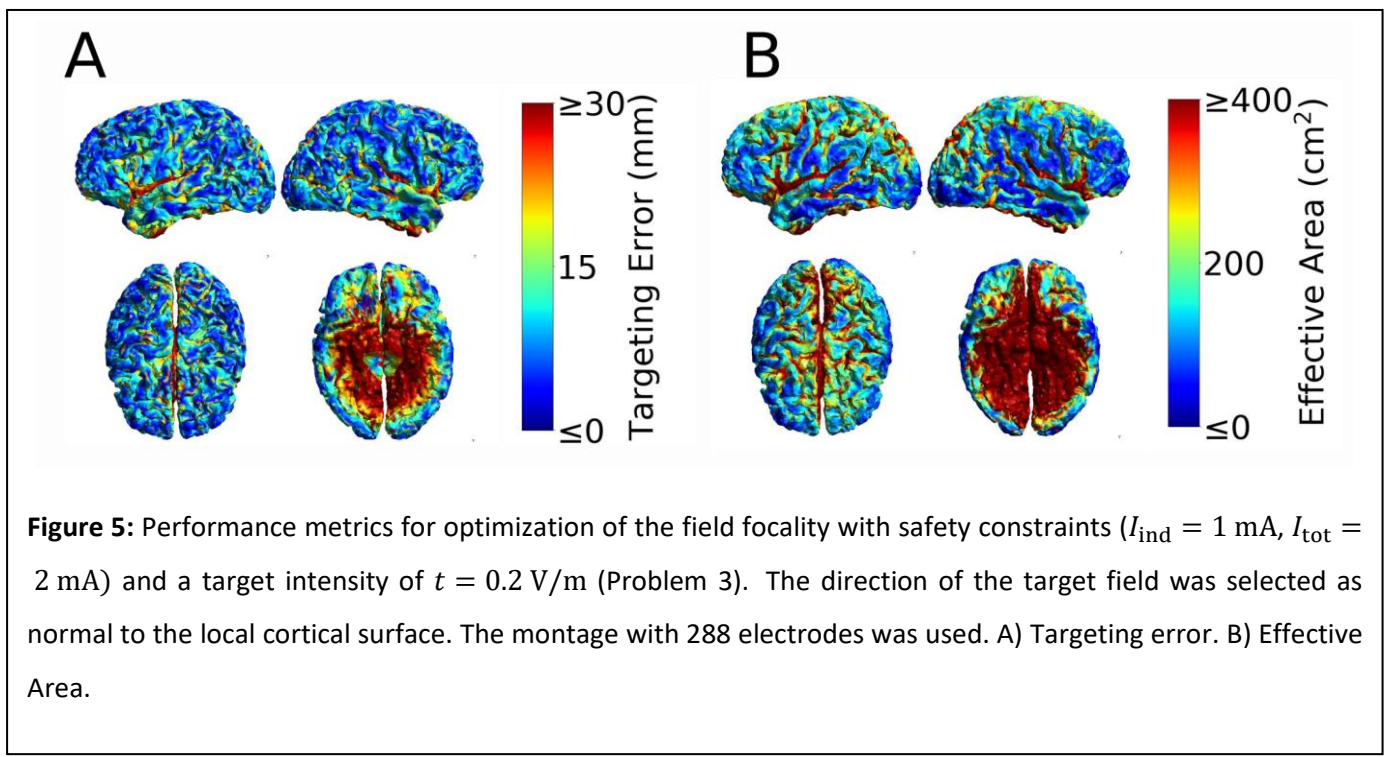

**Figure 5:** Performance metrics for optimization of the field focality with safety constraints ($I_{\text{ind}} = 1$ mA, $I_{\text{tot}} = 2$ mA) and a target intensity of $t = 0.2$ V/m (Problem 3). The direction of the target field was selected as normal to the local cortical surface. The montage with 288 electrodes was used. A) Targeting error. B) Effective Area.

Without safety constraints in place, the optimization algorithm minimizes the overall electric field in the cortex by mainly two strategies: (1) Selecting nearby electrodes in order to increase the focality of the field in the cortex. However, this causes extensive shunting, and therefore high current is needed. (2) Cancelling out electric fields outside the target region by injecting small currents through many electrodes, so that their fields mutually cancel out in non-target regions. However, limiting the total amount of injected currents limits the usage of close-by electrodes, which are ineffective due to the high degree of shunting, and strongly reduces the extent to which compensatory currents can be applied.

Expectedly, this effect is more pronounced for higher target fields, where most or all of the allowed current is fed into the head through a few electrodes in order to achieve the required field in the target. In turn, little current is left that can be used to create cancelation fields (please see Suppl. Fig. S8 for a demonstration of this effect for target electric fields varied between 0.1 V/m, 0.2 V/m and 0.3 V/m). This results in a trade-off between intensity at the target and focality, as it was also reported by (Dmochowski et al., 2017a; Dmochowski et al., 2011).

In addition to the inferior and medial parts of the brain, regions around the longitudinal fissure are affected most by the safety constraints. Due to the high conductivity of CSF, the currents in brain areas close to the longitudinal fissure tend to point preferentially towards or away from the fissure. For optimization without safety constraints and with a low target intensity, this effect is counteracted by invoking many electrodes, thereby cancelling undesired field directions. However, limiting the totally induced current reduces the effectiveness of this cancellation approach.

Optimizing with safety constraints resulted in spurious small improvements in the Targeting Error at some cortical positions compared to the case without constraints (blue dots in Fig. 5B). This results from the fact that the targeting error is not directly assessed during the optimization procedure. In contrast, the effectively stimulated area is closely related to the delivered energy. As the latter is minimized by the optimization approach, adding safety constraints results in a consistent increase in the Effective Area for all positions.

### Effect of Constraining the Number of Active Electrodes

To evaluate the effect of constraining the number of active electrodes (Problem 4) on the electric field focality, we started by evaluating different approaches to reduce the number of active electrodes. We selected the safety constraints $I_{\text{ind}} = 1$ mA, $I_{tot} = 2$ mA and target field strength of $t = 0.2 \, V/m$, because the results in Figure 4 and 5 showed that this value could still be achieved with relative ease at most cortical positions for the selected safety constraints, but was high enough to make the electrode selections challenging. We tested the performance of three approaches, termed "Full B-B", "Selected B-B" and "Projection" in constraining the active electrodes to $N = 4$. The "Full B-B" approach applies the Branch-and-Bound algorithm (for details, see description in the Methods and in supplementary chapter S.1) on all 288 electrodes. The "Selected B-B" preselects a subset of potentially relevant electrodes using an initial optimization with $I_{tot\_pre} = 3I_{tot} = 6$ mA, and applies the Branch-and-Bound algorithm to this subset only. By that, the computationally more expensive Branch-and-Bound algorithm is applied to fewer electrodes, making the convergence faster. The "Projection" approach runs an optimization with safety constraints, selects the $N$ electrodes with highest current, and runs another optimization involving only those electrodes.

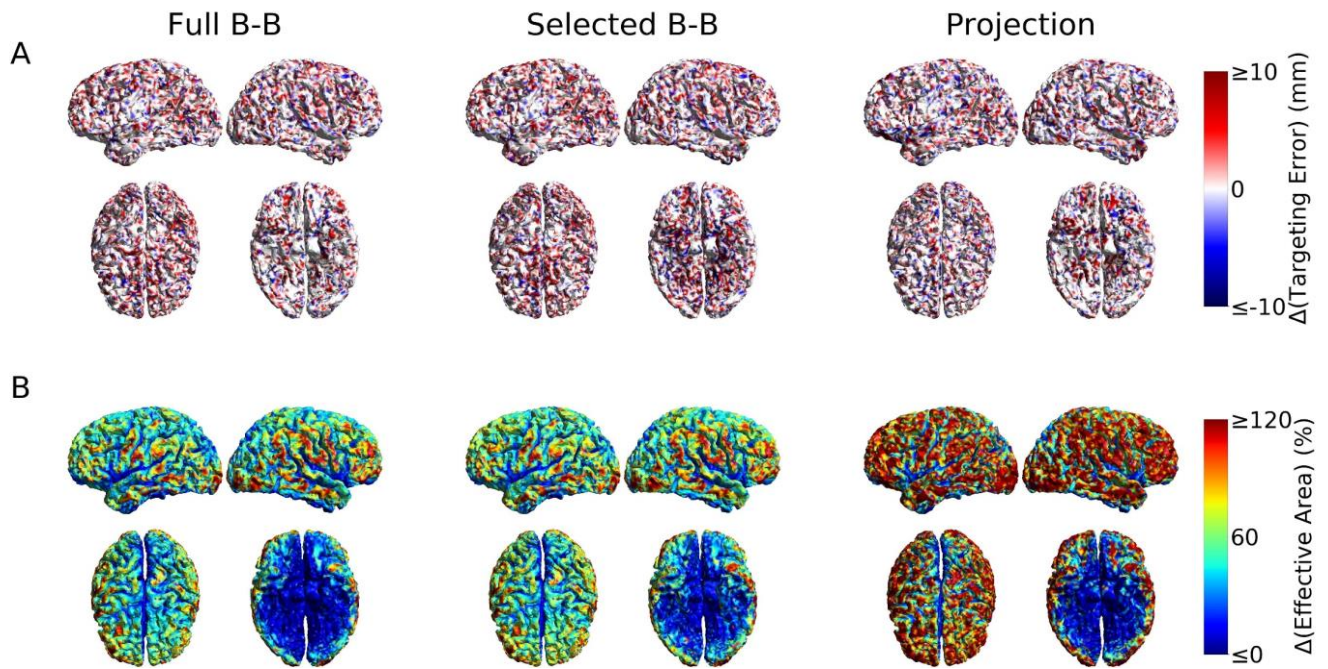

**Figure 6:** Comparison of three methods to restrict the number of active electrodes during optimization of the field focality in the brain (Problem 4). The optimization parameters were $I_{\text{ind}} = 1$ mA, $I_{\text{tot}} = 2$ mA, $t = 0.2$ V/m, and the number of active electrodes was exemplarily restricted to $N = 4$. A) Difference in the achieved Targeting Error, assessed relative to the optimizations without restrictions of the number of active electrodes, but otherwise identical settings (Fig. 5A). B) Difference in the Effective Area, relative to the corresponding optimizations without restrictions of the number of active electrodes (Fig. 5B). Left columns in A and B: Branch-and-Bound algorithm performed on all electrodes (Full B-B). Middle columns: Branch-and-Bound algorithm performed on a pre-selected set of electrodes (Selected B-B). The pre-selection was performed by solving an optimization problem with relaxed safety constraints ($I_{\text{tot}} \leftarrow 3I_{\text{tot}}$), thereby locating electrodes that might potentially contribute to the final optimized montage. Right columns: The "Projection" approach runs an optimization with safety constraints, selects the $N$ electrodes with highest current, and runs another optimization involving only those electrodes.

Generally, restricting the number of active electrodes tends to increase the targeting errors (Fig. 6A & 6B shows difference plots relative to the optimizations with safety constraints depicted in Fig. 5A & 5B) and decreases the focality particularly at gyral crowns. While the three approaches perform similar with respect to minimizing the targeting error, the "Projection" approach is clearly inferior to the two Branch-and-Bound algorithms in maintaining a focal stimulation when compared to the optimizations without restricted number of active electrodes (Fig. 6B). The two Branch-and-Bound algorithms perform very similar, so that we used the faster "Selected B-B" algorithm for the subsequent optimizations with restricted number of electrodes.

These results imply that our method for electrode selection outperforms the simpler "Projection" method, and that the "Selected B-B" heuristic performs very similar to the "Full-BB" approach, albeit being faster.

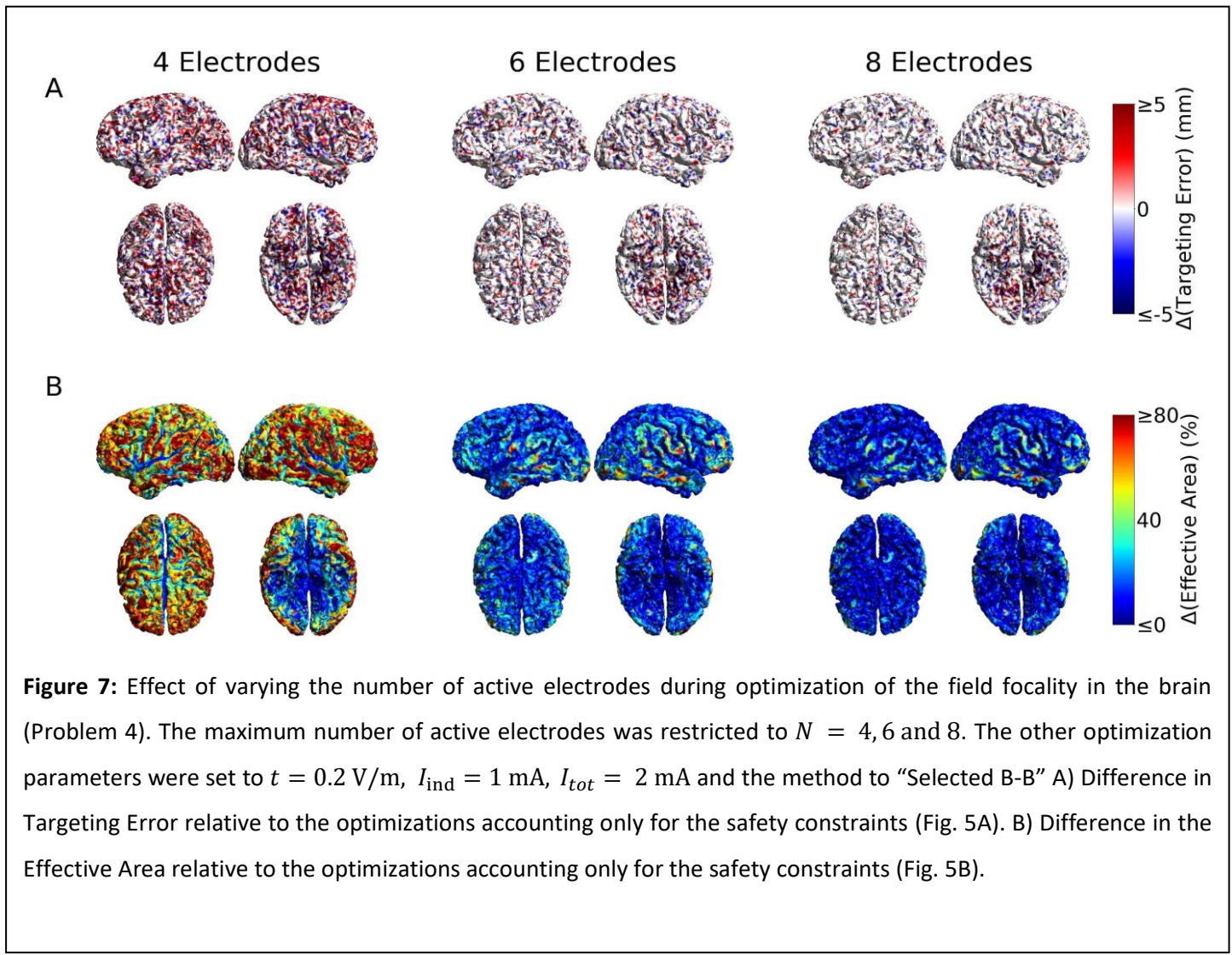

**Figure 7:** Effect of varying the number of active electrodes during optimization of the field focality in the brain (Problem 4). The maximum number of active electrodes was restricted to $N = 4, 6$ and $8$. The other optimization parameters were set to $t = 0.2$ V/m, $I_{\text{ind}} = 1$ mA, $I_{tot} = 2$ mA and the method to "Selected B-B" A) Difference in Targeting Error relative to the optimizations accounting only for the safety constraints (Fig. 5A). B) Difference in the Effective Area relative to the optimizations accounting only for the safety constraints (Fig. 5B).

We evaluated the effect of limiting the number of electrodes on the achieved field focality (Problem 4). Figure 7 shows the effect of limiting the number of active electrodes $N$ to 4, 6 and 8. The other parameters were set to $t = 0.2$ V/m, $I_{\text{ind}} = 1$ mA, $I_{tot} = 2$ mA and "Selected B-B". The figure depicts a comparison with the field metrics obtained with and without constraining the number of electrodes (Fig. 5A & 5B) for normal field orientations. Allowing for only four active electrodes seems to produce less focal electric fields in most of the brain. However, the electric field focality seem to improve strongly when allowing for six electrodes to be active. Further improvements, mainly in the gyral crowns, can be achieved using eight active electrodes. Increasing the number of electrodes beyond eight improves the results only marginally further. Suppl. Figures S11 and S12 show the corresponding results for subcortical targets and tangential target orientations, respectively, confirming the above results. Taken together, even though a dense electrode grid is needed to ensure that the best electrode positions can be chosen by the optimization procedure, our results indicate that users of multi-channel stimulators benefit little from using more than 8 active electrodes for most cortical targets.

It is interesting to notice that the electric field in inferior positions is not affected by limiting the number of electrodes to four. This is because 0.2 V/m is close to the maximum electric field in these positions (see Figure 4). As pointed out in the *Methods*, the maximum electric field is achieved with $\text{ceil}(2I_{tot}/I_{ind}) = 4$ electrodes, meaning that the maximum number of active electrodes in these positions is naturally limited to four.

### Effect of Constraining the Angular Deviation of the Electric Field in the Target

We evaluate the effect of explicitly constraining the angle of the electric field in the target region, in addition to obeying safety constraints (Problem 6). Figure 8 shows the results for constraining the maximum angle to $\theta_{\max} = 15°, \ 22.5°$ and $30°$ in relation to the desired target field direction that was chosen to be normal to the cortical sheet. The other optimization parameters were set to $t = 0.2$ V/m, $I_{\text{ind}} = 1$ mA and $I_{tot} = 2$ mA. The results are shown as difference plots relative to the results depicted in Fig. 5A & 5B.

Interestingly, the angle constraint causes only a slight reduction in the achievable targeting accuracy and field focality in most brain areas, except for regions that are above the low conductive skull base and are surrounded by large amounts of CSF (Fig. 1A). At the latter positions, the electric fields exhibit a profound tendency to point in the lateral direction towards CSF, so that targeting accuracy and field focality are compromised by enforcing a field direction normal to the cortical sheet. Suppl. Figs. S13 and S14 show the same analysis for subcortical targets and for targets with tangent orientations to the cortical sheet, respectively, confirming the above results.

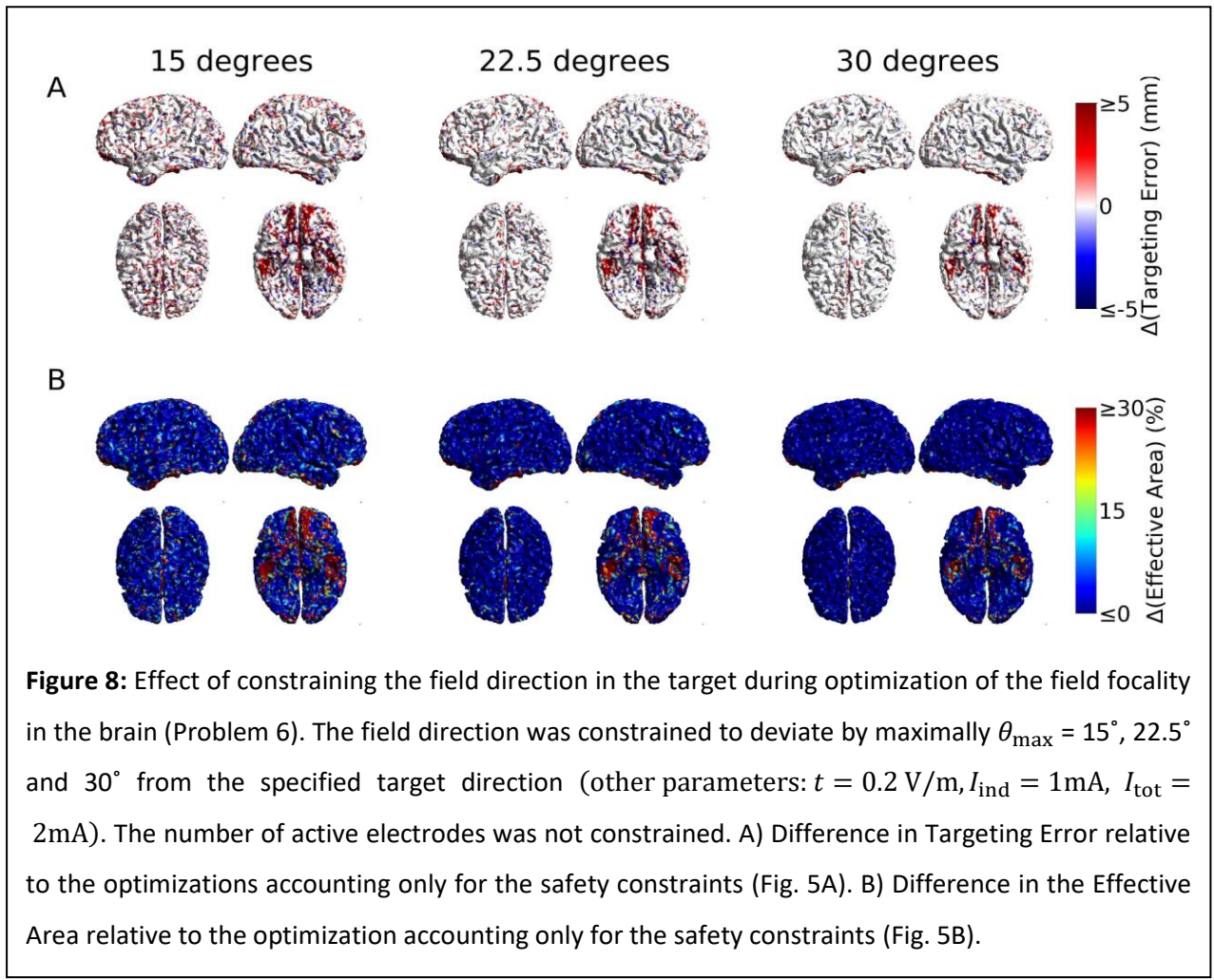

**Figure 8:** Effect of constraining the field direction in the target during optimization of the field focality in the brain (Problem 6). The field direction was constrained to deviate by maximally $\theta_{max}$ = 15°, 22.5° and 30° from the specified target direction (other parameters: $t = 0.2$ V/m, $I_{ind} = 1$ mA, $I_{tot} = 2$ mA). The number of active electrodes was not constrained. A) Difference in Targeting Error relative to the optimizations accounting only for the safety constraints (Fig. 5A). B) Difference in the Effective Area relative to the optimization accounting only for the safety constraints (Fig. 5B).

Constraining also the number of active electrodes, in addition to constraining the angle of the electric field in the target region and obeying to the safety constraints, can be done with low additional cost with regards to the achievable targeting accuracy and field focality (Problem 7). As an example, Figure 9 shows the results for constraining number of electrodes to N = 4, 6 and 8 and the field direction in the target to $\theta_{max} = 22.5°$. The other optimization parameters were set to $t = 0.2$V/m, $I_{ind} = 1$ mA, $I_{tot} = 2$ mA and "Selected B-B". The results are shown as difference plots relative to the results depicted in Fig. 5A & 5B.

The similarity between the results in Figure 9 and Figure 7 demonstrates that ensuring that the field only deviates slightly from the desired field direction can be done at a small additional cost and confirms that a relatively small number of active electrodes (~8) is still sufficient in that case. Suppl. Figs. S15 and S16 show the same analysis for subcortical targets and for targets with tangent orientations to the cortical sheet, respectively, confirming the above results.

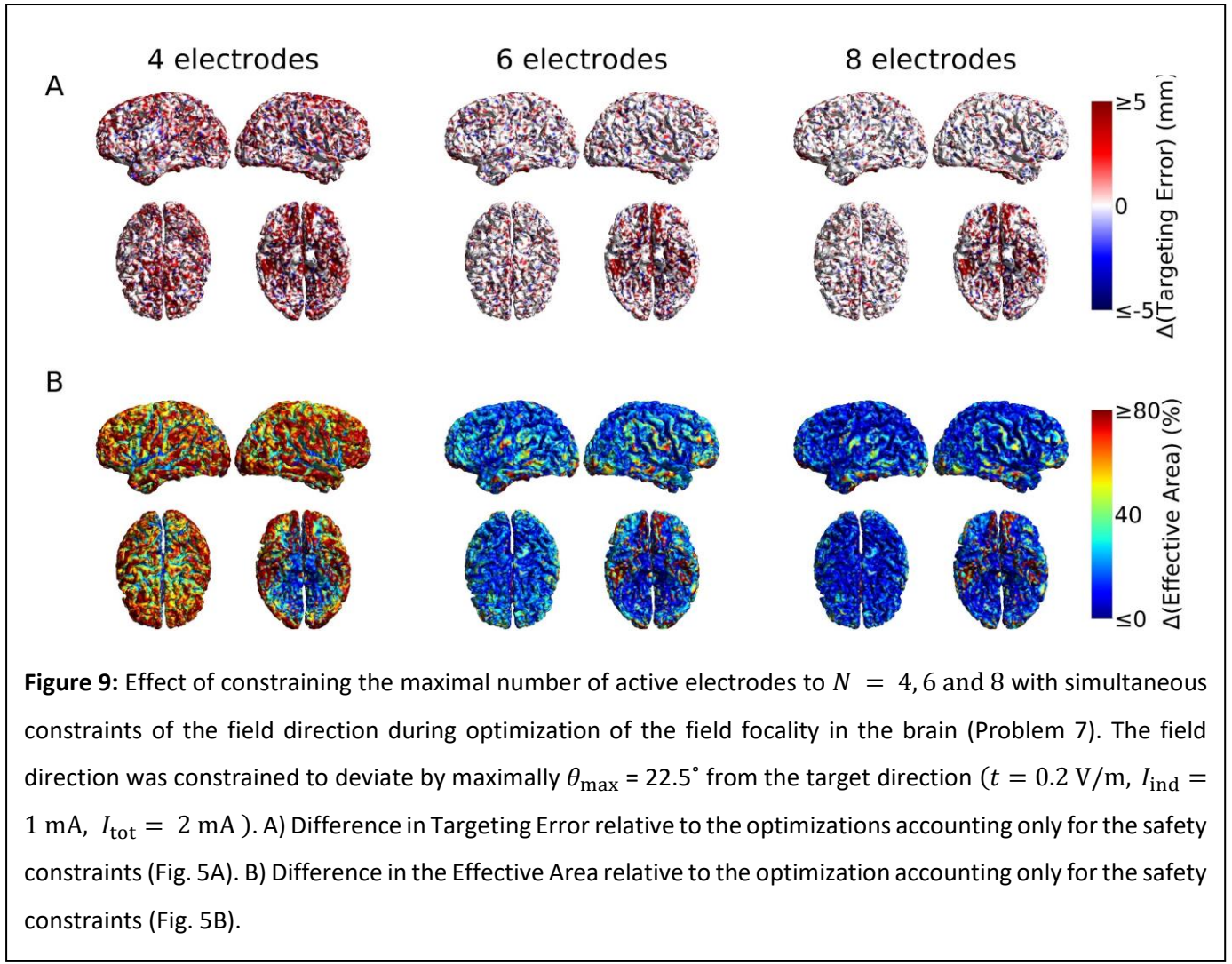

**Figure 9:** Effect of constraining the maximal number of active electrodes to $N = 4, 6$ and 8 with simultaneous constraints of the field direction during optimization of the field focality in the brain (Problem 7). The field direction was constrained to deviate by maximally $\theta_{max}$ = 22.5° from the target direction ($t = 0.2$ V/m, $I_{ind} = 1$ mA, $I_{tot} = 2$ mA). A) Difference in Targeting Error relative to the optimizations accounting only for the safety constraints (Fig. 5A). B) Difference in the Effective Area relative to the optimization accounting only for the safety constraints (Fig. 5B).

## Multi-Target Optimization

To demonstrate the performance of our approach for multi-target optimization (Problem 10), we tested a configuration where the gyral crowns of the left and right hand knob areas of the motor cortices where simultaneously targeted (N = 4 to 10 in steps of 2, $t = 0.2$V/m in both targets, $I_{ind} = 1$ mA, $I_{tot} = 2$ mA, "Selected B-B"). The field direction was chosen normal to the local cortical surface at both targets. Two sets of optimizations were performed, the first with the field directions pointing outwards at both targets (out/out) and an additional configuration where the local field directions were opposite (in/out).

The resulting electric fields are shown in Figure 10, confirming that the fields are indeed focused around the two targets. The two electric field hotspots become more confined and are gradually decoupled from each other as the number of active electrodes increases from 4 to 8. Further increasing to 10 active electrodes gives little additional improvement. This is confirmed by the metrics for the targeting accuracy and field focality (Suppl. Table 1) for the fields shown in Figure 10. Taken together, our formulation of the optimization problem allows for a balanced simultaneous targeting of multiple brain regions. Even in the case of two targets, we do not see large benefits of using more than 8 active electrodes. However, only two target combinations were tested here, so that is unclear whether this observation holds in general.

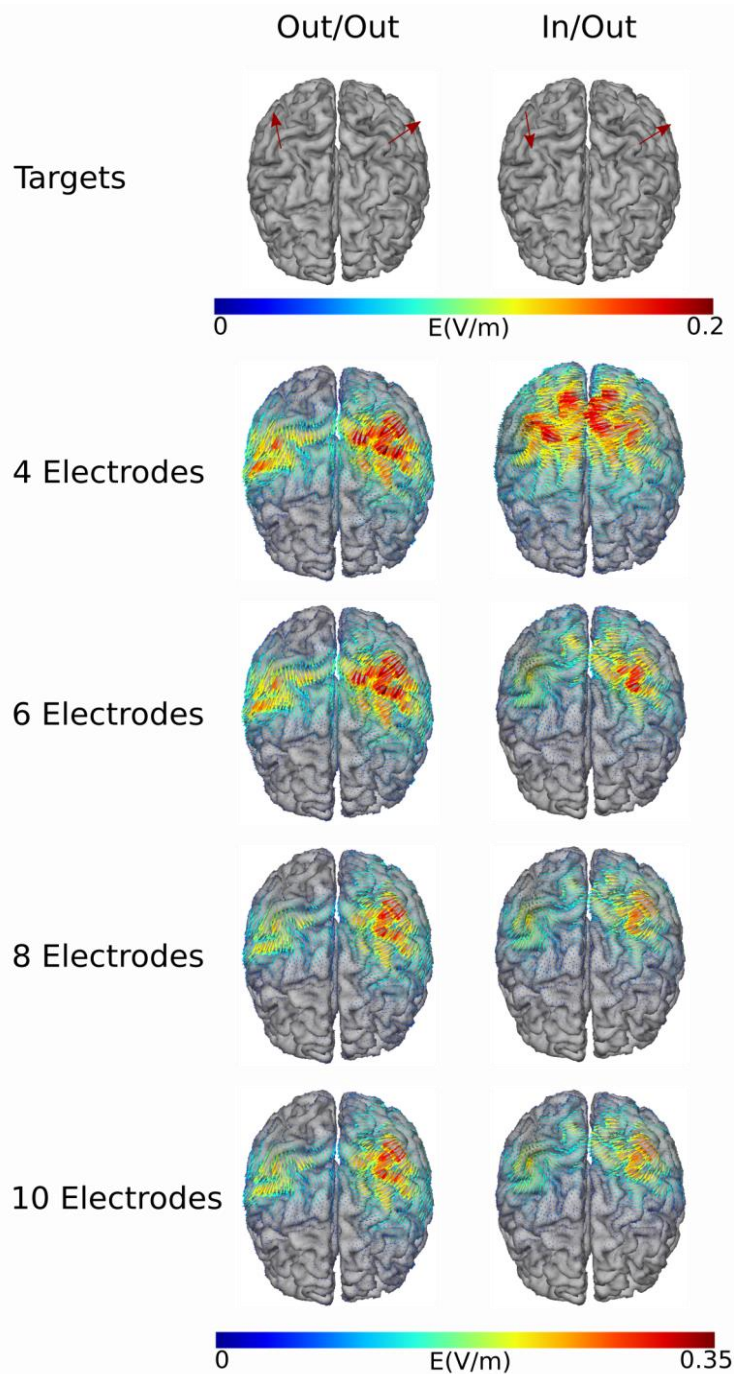

**Figure 10:** Electric field distributions with optimized focality for simultaneous targeting of the left and right handknob areas of the motor cortices (Problem 10). The constraints were set to ($t = 0.2$ V/m in each of the two targets, $I_{\text{ind}} = 1$ mA, $I_{\text{tot}} = 2$ mA), and the number of electrodes was constrained to $N = 4, 6, 8$ and $10$. In the left column, we set the fields in both targets to point outwards, and in the right column, we set the field in the right motor cortex to point inward and the field in the left motor cortex to point outwards.

## Time to run Optimizations

In order to evaluate the time required for the optimizations with our algorithms, we assessed the times for solving Problems 2, 3, 4, 6 and 7 with safety constraints $I_{ind} = 1$ mA and $I_{tot} = 2$ mA. For all except Problem 2 (that maximizes the intensity), a target intensity of $t = 0.2$ V/m was used. In addition, the number of active electrodes was constrained to $N = 6$ using the "Selected B-B" algorithm and the angle deviation was constrained to $\theta_{max}$ = 22.5°, if required by the tested problem. We used the lead-field defined on the simplified middle grey matter model (20,000 nodes) and with $n = 288$ electrodes. We recorded the time taken to calculate the $Q$, $l$ and $Q_t$ matrices and perform the optimizations a random sample of 1,000 positions with targets defined in a direction normal to the cortical surface on a Lenovo ThinkPad laptop computer with an Intel i7-7500U processor (2 cores, 4 threads), 16 GB of memory, a SSD hard-drive and running Ubuntu Linux 18.04.

| Problem # | Min. time (s) | Median time (s) | Max. time (s) |
|---|---|---|---|
| 2 | 0.48 | 0.49 | 0.72 |
| 3 | 0.49 | 0.68 | 1.15 |
| 4 | 0.68 | 1.39 | 9.14 |
| 6 | 0.47 | 0.94 | 3.60 |
| 7 | 0.77 | 3.82 | 138.6 |

**Table 3:** Time to run the optimization algorithm under various settings. To acquire the timing, we ran optimizations on 1000 randomly selected points on the subsampled grey matter surface, using the leadfield created with 288

Table 3 shows the minimum, median and maximum time to run the optimizations. Problem 2 (intensity maximization) is the simplest problem to solve as it is a linear programming problem, which is reflected by the timing results. Problem 3 (optimization of field focality) is a more complex quadratic programming problem, but is still solved in less than a second in most cases by our implementation. Problem 4 (optimization of field focality with constraints on the number of active electrodes) is not a convex problem anymore. However, the timings show that the implemented Branch-and-Bound algorithm handles the L0 constraint efficiently, as the median times are only two times larger than the ones observed in the problem without constraint on the number of active electrodes. Problem 6 (optimization of field focality with constraints on the field angle in the target) is solved in a sub-optimal way, as a conic solver would be more appropriate, but still requires less than one second in most cases. Problem 7 (optimization of field focality with constraints on the field angle in the target and the number of active electrodes) is the most demanding case but is still solved in a few seconds on average and a few minutes in worst case.

# DISCUSSION AND CONCLUSION

## Summary of the algorithm and implementation

Several and in part opposing objectives exist for the optimization of multi-electrode TES montages (e.g., intensity at the target vs. field focality) and the optimization results are further influenced by safety and technical constraints (e.g., limits on the maximally injected current and maximal number of active electrodes). These different combinations of objectives and constraints give rise to a variety of optimization problems, which have different solutions and often require different algorithms. In the current paper, we consider several practically relevant optimization problems for multi-electrode TES montages and use a common mathematical framework to sort these problems according to their characteristics and complexity.

Unless the aim is to maximize the field in the target (Problems 2, 5 and 9), most of the considered problems share the common objective to minimize the field energy in the brain in order to obtain focal fields. In our implementation, the user sets them up by defining the desired strength and direction of the electric field in the target region, which are practically meaningful and intuitive parameters. All considered problems, except Problem 1, which is a theoretical best-case scenario, obey safety limits, which are defined by upper limits of the currents injected per electrode and in total. When desired, also the maximal number of active electrodes can be set to match the number of available stimulation channels, and fine control of the maximal angular deviation of the target field from the desired direction can be established in addition.

Our approach shares several features with previously published work. For example, Guler et al. (2016a) proposed an optimization problem that is based on similar quantities as used in the current work (field at target, energy and current injection limits). However, in his approach, the maximum energy is selected by the user while the field in the target region and direction is maximized. We believe that our approach has the advantage of using a more intuitive parameter as user input, as electric field values that can serve as reference are often reported in simulation studies, while energy values are rarely stated. Dmochowski et al. (2011) proposed two types of optimization problems, a least squares approach and a linearly constrained minimum variance approach (LCMV). The least squares approach is based on the selection of a target region and field direction, followed by a manual tuning of a weight parameter until an acceptable field intensity is achieved in the target. The LCMV approach closely resembles our optimization of the field focality as done in Problem 2. The main difference is that in the LCMV approach defined by Dmochowski et al. all electric field components in the complete target region need to be specified and fulfilled exactly, while we use a more relaxed constraint, in which only the average strength of the field in the target along the specified direction needs to be fulfilled. As our approach leaves more degrees of freedom for the solution, we expect to obtain more focal fields, albeit at the cost of less control of the field at the target. Other approaches use different quantities, such as Wagner et al. (2015) where the user selects a maximum field value outside the target region, or (Ruffini et al., 2014) where a similarity measure to a target field is optimized.

We addressed the practically important issue of limiting the number of electrodes, given that TES stimulators usually have a small number of channels available. This was done in a principled and efficient way by using Branch-and-Bound algorithms. Noticeably, this algorithm does not only promote quick convergence, but also certifies that the obtained solution is close to the global optimum. Branch-and-bound algorithms were already considered for a similar problem (Guler et al., 2016b), but with substantially lower performance requiring a few hours to converge for a single simulation. The higher efficiency of our implementation (Table 3) is probably due to more suitable upper and lower bound functions and the state-space definition. In addition, we established and evaluated a useful heuristic for further accelerating the Branch-and-Bound algorithm by preselecting "candidate" electrodes (Figs. 6 & 7). Competing algorithms such as Genetic Algorithms (Ruffini et al., 2014) provide no guarantee of convergence towards or proximity to the global optima, and are also reported to need in the order of a few hours to converge.

To the best of our knowledge, this is the first optimization study that explicitly controls the electric field angle in the target, besides the LCMV approach by (Dmochowski et al., 2011), where the angle needs to be reached exactly. Our implementation was successful in solving the underlying problem (a QCQP), and reasonably efficient obtaining results in less than one second in most cases. The implementation is likely less efficient than a cone solver, but it has the advantage of keeping the code base simpler due to its similarity to the other considred problems. This new procedure will allow for more in-depth investigations of the effect of angle on TES outcome.

## Summary of findings

The efficiency of the optimization algorithms, together with the practically meaningful and intuitive input parameters in our optimizations, allowed us to perform an extensive mapping of the optimizations and test the effect of various constraints and parameter choices across the entire cortex. Our results demonstrate that local anatomy plays a significant role, as a far better field focality, better targeting accuracy and higher intensities can be obtained for targets in superficial brain areas compared to sulcal or deep targets (Figs. 3 & 4). This is the case even for a theoretical best-case scenario without safety constraints. Compared to this scenario, adding safety constraints strongly reduces the obtainable field focality and targeting accuracy further, in particular for deeper targets (Fig. 5). These observations are also in line with the intensity-focality trade-off reported in (Dmochowski et al., 2011; see our Suppl. Figures S8-S10). Finally, additionally constraining the angle of the field in the target to meet the desired direction has a low additional penalty for most target positions (Fig. 8). While the results in the main paper were obtained for cortical targets being oriented normally to the gray matter sheet, we confirmed that the findings generalize to tangential target directions and to targets in deep areas (Suppl. Figs. S1-S16).

It seems beneficial to perform the optimization of multi-electrode TES montages on a fine electrode grid (up to ~300 electrodes, Suppl. Fig. S2-S7), so that the optimization procedure can do fine adjustments of the positions of the active electrodes in order to obtain the most focal and intense electric fields. This is particularly relevant when aiming to focus the field on superficial parts of the brain. However, it is important to note that this fine grid is only required for the planning of the optimized montage using the virtual head model, while only a few active electrodes are finally required for the stimulation. However, this suggests that the positions of these electrodes are required to be accurately controlled in practice.

We observed that constraints on the total current delivered (sum of absolute electrode currents) severely limits the benefit of increasing the number of active electrodes beyond ~8, at least when targeting a single area. This happens because the limits on the total current also limit the ability to recruit secondary electrodes for cancelling the field in positions of undesired stimulation, and therefore has effect similar to limiting the total number of electrodes. A counterintuitive consequence of this is that a limitation of the total current actually causes more energy to be delivered to the brain, given that the target field intensity can be met. Therefore, in order to archive the full benefit of focal optimized multi-electrode stimulation, it would be useful to consider an extension of the implemented safety constraints and possibly rather ensure that the maximum current density in relevant tissues classes are kept below a safety limit. In particular, limiting the current density in the skin to avoid discomfort and heating of tissue due to stimulation not only immediately beneath, but also in between the electrodes might be a rationale choice for maximizing the benefit of focal optimized multi-electrode TES.

Our implementation of TES multi-electrode optimization also allows for the incorporation of multiple targets. We exemplarily demonstrated that the algorithm successfully created balanced focal fields around two distant target regions, while also keeping the number of active electrodes at a reasonably low number (Fig. 10). For targets placed closely together, we might not be able to fully decouple the electric fields around each target, in contrast to the situation tested in Figure 10. This happens as the electrodes involved in the stimulation of one target cause a significant electric field also in the other target, as observed previously in (Saturnino et al., 2017), so that a stimulation of the region between the two targets cannot be fully avoided.

## Limitations

Even though we explored a large range of parameters, some were kept constant during all tests, most importantly the head model and the safety constraints. Inter-individual variations in anatomy can impact on the exact values for the optimum field focality, especially in the case of disease (Dmochowski et al., 2013). However, we do not expect to reach fundamentally different conclusions on the influence of the optimization parameters or constraints when changing the head model. Even though specific anatomical features might enhance or counteract the effect of, for example, adding safety constraints or limiting the number of electrodes, these effects stem from the underlying physics involved in TES generated electric fields. Changing the safety constraints may have an effect on the focality-intensity trade-off (Fig. 5) and on the number of electrodes needed for a focal stimulation (Figs. 7 & 9). If we increase the total current bound $I_{\text{tot}}$, we would expect the trade-off to persist, yet to be less pronounced, because we more of the total current budget can be used then to cancel out electric fields outside the target area. Decreasing the individual current bound $I_{\text{ind}}$ while keeping the total current bound $I_{\text{tot}}$ constant would result in more active electrodes and likely less focal fields, as electrodes in sub-optimal positions would be needed to reach the desired electric field at the target.

Importantly, there are many other factors besides variations in the strength and spatial distribution of the electric field that can cause variability in the outcome of an TES experiment, such as the brain state, age, genetics and neurochemistry (Karabanov et al., 2016; Ridding and Ziemann, 2010), which are not addressed within the optimization framework presented in the current work.

## Conclusion

We introduce a new algorithm to optimize TES electric fields and assess the impact of optimization parameters and constraints. We show that the maximally achievable field focality is fundamentally limited by the anatomy of the head and its physical properties, and demonstrate a focality-intensity trade-off. Given commonly implemented safety constraints, we show that there is little benefit in using more than eight electrodes for TES electrode montages, even though the optimization benefits of using a fine electrode grid for planning the electrode montage. Furthermore, we found that the electric field can be controlled with little penalty on field focality and that multiple distant targets can be optimized while keeping effects of the stimulation balanced. The optimization code used for this study will be available in a future version of our open source software SimNIBS. This study was limited to traditional TES methods. More recent methods, such as Temporally Interfering (TI) Electric Fields (Grossman et al., 2017) might be able to generate more focal fields in deeper brain areas than traditional TES methods. Algorithms to optimize TI fields and study their limits and relationship to optimization parameters and constraints are an interesting topic for future research.

## ACKNOWLEDGEMENTS

This study was supported by the Lundbeck foundation (grant no. R244-2017-196 and R118-A11308), and the Novo Nordisk foundation (grant no. NNF14OC0011413). Hartwig R. Siebner holds a 5-year professorship in precision medicine at the Faculty of Health Sciences and Medicine, University of Copenhagen which is sponsored by the Lundbeck Foundation (Grant Nr. R186-2015-2138).